\pdfoutput=1
\documentclass[12pt,a4paper]{article}

\usepackage{ifthen}
\newboolean{pdflatex}
\setboolean{pdflatex}{true}
\usepackage{tikz-feynman}
\newboolean{articletitles}
\setboolean{articletitles}{true}
\newboolean{uprightparticles}
\setboolean{uprightparticles}{false}
\newboolean{inbibliography}
\setboolean{inbibliography}{false}

\def\paperauthors{LHCb collaboration}
\def\paperasciititle{Measurements of the branching fractions of Lambda_c+ -> p pi- pi+, Lambda_c+ -> p K- K+, and Lambda_c+ -> p pi- K+}
\def\papertitle{Measurements of the branching fractions of \LcpToppimpip, \LcpTopKmKp, and \LcpToppimKp}
\def\paperkeywords{{High Energy Physics}, {LHCb}}
\def\papercopyright{CERN on behalf of the LHCb collaboration}
\def\paperlicence{CC-BY-4.0}
\def\paperlicenceurl{https://creativecommons.org/licenses/by/4.0/}
\usepackage[top=1in, bottom=1.25in, left=1in, right=1in]{geometry}
\usepackage{microtype}
\usepackage{lineno}  
\usepackage{xspace} 
\usepackage{caption}

\usepackage{graphicx}
\usepackage{color}
\usepackage{colortbl}
\graphicspath{{./figs/}}

\usepackage{amsmath}
\usepackage{amssymb}
\usepackage{amsfonts}
\usepackage{upgreek}

\usepackage{subfig}
\usepackage{multirow}

\newcommand*\patchAmsMathEnvironmentForLineno[1]{%
\expandafter\let\csname old#1\expandafter\endcsname\csname #1\endcsname
\expandafter\let\csname oldend#1\expandafter\endcsname\csname
end#1\endcsname
 \renewenvironment{#1}%
   {\linenomath\csname old#1\endcsname}%
   {\csname oldend#1\endcsname\endlinenomath}%
}
\newcommand*\patchBothAmsMathEnvironmentsForLineno[1]{%
  \patchAmsMathEnvironmentForLineno{#1}%
  \patchAmsMathEnvironmentForLineno{#1*}%
}
\AtBeginDocument{%
\patchBothAmsMathEnvironmentsForLineno{equation}%
\patchBothAmsMathEnvironmentsForLineno{align}%
\patchBothAmsMathEnvironmentsForLineno{flalign}%
\patchBothAmsMathEnvironmentsForLineno{alignat}%
\patchBothAmsMathEnvironmentsForLineno{gather}%
\patchBothAmsMathEnvironmentsForLineno{multline}%
\patchBothAmsMathEnvironmentsForLineno{eqnarray}%
}

\usepackage{hyperxmp}

\usepackage[pdftex,
            pdfauthor={\paperauthors},
            pdftitle={\paperasciititle},
            pdfkeywords={\paperkeywords},
            pdfcopyright={Copyright (C) \papercopyright},
            pdflicenseurl={\paperlicenceurl}]{hyperref}

\usepackage[all]{hypcap}

\usepackage{xspace} 
\usepackage{upgreek}

\def\lhcb {\mbox{LHCb}\xspace}

\def\belle  {\mbox{Belle}\xspace}

\def\lhc    {\mbox{LHC}\xspace}

\def\MagUp {\mbox{\em Mag\kern -0.05em Up}\xspace}

\ifthenelse{\boolean{uprightparticles}}%
{\def\Palpha      {\ensuremath{\upalpha}\xspace}

 \def\Pmu         {\ensuremath{\upmu}\xspace}                 
 \def\Pnu         {\ensuremath{\upnu}\xspace}                 
 \def\Pxi         {\ensuremath{\upxi}\xspace}                 
 \def\Ppi         {\ensuremath{\uppi}\xspace}                 
                  
 \def\Prho        {\ensuremath{\uprho}\xspace}

 \def\PDelta      {\ensuremath{\Delta}\xspace}                 
 \def\PXi      {\ensuremath{\Xi}\xspace}                 
 \def\PLambda      {\ensuremath{\Lambda}\xspace}                 
 \def\PSigma      {\ensuremath{\Sigma}\xspace}                 
 \def\POmega      {\ensuremath{\Omega}\xspace}                 
 \def\PUpsilon      {\ensuremath{\Upsilon}\xspace}

 \def\PB      {\ensuremath{\mathrm{B}}\xspace}                 
                  
 \def\PD      {\ensuremath{\mathrm{D}}\xspace}

 \def\PK      {\ensuremath{\mathrm{K}}\xspace}

 \def\PW      {\ensuremath{\mathrm{W}}\xspace}

 \def\Pb      {\ensuremath{\mathrm{b}}\xspace}                 
 \def\Pc      {\ensuremath{\mathrm{c}}\xspace}                 
 \def\Pd      {\ensuremath{\mathrm{d}}\xspace}

 \def\Ph      {\ensuremath{\mathrm{h}}\xspace}                 
 \def\Pi      {\ensuremath{\mathrm{i}}\xspace}

 \def\Pm      {\ensuremath{\mathrm{m}}\xspace}

 \def\Pp      {\ensuremath{\mathrm{p}}\xspace}                 
                  
 \def\Pr      {\ensuremath{\mathrm{r}}\xspace}                 
 \def\Ps      {\ensuremath{\mathrm{s}}\xspace}                 
                  
 \def\Pu      {\ensuremath{\mathrm{u}}\xspace}

 \def\Px      {\ensuremath{\mathrm{x}}\xspace}                 
                  
 \def\Pz      {\ensuremath{\mathrm{z}}\xspace}                 
}
{\def\Palpha      {\ensuremath{\alpha}\xspace}

 \def\Pmu         {\ensuremath{\mu}\xspace}                 
 \def\Pnu         {\ensuremath{\nu}\xspace}                 
 \def\Pxi         {\ensuremath{\xi}\xspace}                 
 \def\Ppi         {\ensuremath{\pi}\xspace}                 
                  
 \def\Prho        {\ensuremath{\rho}\xspace}

 \mathchardef\PDelta="7101
 \mathchardef\PXi="7104
 \mathchardef\PLambda="7103
 \mathchardef\PSigma="7106
 \mathchardef\POmega="710A
 \mathchardef\PUpsilon="7107
                  
 \def\PB      {\ensuremath{B}\xspace}                 
                  
 \def\PD      {\ensuremath{D}\xspace}

 \def\PK      {\ensuremath{K}\xspace}

 \def\PW      {\ensuremath{W}\xspace}

 \def\Pb      {\ensuremath{b}\xspace}                 
 \def\Pc      {\ensuremath{c}\xspace}                 
 \def\Pd      {\ensuremath{d}\xspace}

 \def\Ph      {\ensuremath{h}\xspace}                 
 \def\Pi      {\ensuremath{i}\xspace}

 \def\Pm      {\ensuremath{m}\xspace}

 \def\Pp      {\ensuremath{p}\xspace}                 
                  
 \def\Pr      {\ensuremath{r}\xspace}                 
 \def\Ps      {\ensuremath{s}\xspace}                 
                  
 \def\Pu      {\ensuremath{u}\xspace}

 \def\Px      {\ensuremath{x}\xspace}                 
                  
 \def\Pz      {\ensuremath{z}\xspace}                 
}

\makeatletter
\ifcase \@ptsize \relax
  \newcommand{\miniscule}{\@setfontsize\miniscule{4}{5}}
\or
  \newcommand{\miniscule}{\@setfontsize\miniscule{5}{6}}
\or
  \newcommand{\miniscule}{\@setfontsize\miniscule{5}{6}}
\fi
\makeatother

\DeclareRobustCommand{\optbar}[1]{\shortstack{{\miniscule (\rule[.5ex]{1.25em}{.18mm})}
  \\ [-.7ex] $#1$}}

\def\mun        {{\ensuremath{\Pmu^-}}\xspace} 

\def\neub       {{\ensuremath{\overline{\Pnu}}}\xspace}

\def\neumb      {{\ensuremath{\neub_\mu}}\xspace}

\def\W      {{\ensuremath{\PW}}\xspace}
\def\Wp     {{\ensuremath{\PW^+}}\xspace}

\def\uquark    {{\ensuremath{\Pu}}\xspace}
\def\uquarkbar {{\ensuremath{\overline \uquark}}\xspace}

\def\dquark    {{\ensuremath{\Pd}}\xspace}
\def\dquarkbar {{\ensuremath{\overline \dquark}}\xspace}

\def\squark    {{\ensuremath{\Ps}}\xspace}
\def\squarkbar {{\ensuremath{\overline \squark}}\xspace}

\def\cquark    {{\ensuremath{\Pc}}\xspace}

\def\bquark    {{\ensuremath{\Pb}}\xspace}

\def\pion   {{\ensuremath{\Ppi}}\xspace}

\def\pip    {{\ensuremath{\pion^+}}\xspace}
\def\pim    {{\ensuremath{\pion^-}}\xspace}

\def\kaon    {{\ensuremath{\PK}}\xspace}
  \def\Kbar    {{\kern 0.2em\overline{\kern -0.2em \PK}{}}\xspace}

\def\KorKbar    {\kern 0.18em\optbar{\kern -0.18em K}{}\xspace}

\def\Kp      {{\ensuremath{\kaon^+}}\xspace}
\def\Km      {{\ensuremath{\kaon^-}}\xspace}

\def\KS      {{\ensuremath{\kaon^0_{\mathrm{ \scriptscriptstyle S}}}}\xspace}

  \def\Dbar    {{\kern 0.2em\overline{\kern -0.2em \PD}{}}\xspace}
\def\D       {{\ensuremath{\PD}}\xspace}

\def\DorDbar    {\kern 0.18em\optbar{\kern -0.18em D}{}\xspace}
\def\Dz      {{\ensuremath{\D^0}}\xspace}

\def\Dp      {{\ensuremath{\D^+}}\xspace}

\def\Dstarp  {{\ensuremath{\D^{*+}}}\xspace}

\def\B       {{\ensuremath{\PB}}\xspace}
\def\Bbar    {{\ensuremath{\kern 0.18em\overline{\kern -0.18em \PB}{}}}\xspace}

\def\BorBbar    {\kern 0.18em\optbar{\kern -0.18em B}{}\xspace}

\def\Y#1S{\ensuremath{\PUpsilon{(#1S)}}\xspace}

\def\proton      {{\ensuremath{\Pp}}\xspace}

\def\Deltares    {{\ensuremath{\PDelta}}\xspace}

\def\Lz          {{\ensuremath{\PLambda}}\xspace}
\def\Lbar        {{\ensuremath{\kern 0.1em\overline{\kern -0.1em\PLambda}}}\xspace}
\def\LorLbar    {\kern 0.18em\optbar{\kern -0.18em \PLambda}{}\xspace}
\def\Lambdares   {{\ensuremath{\PLambda}}\xspace}

\def\Lb      {{\ensuremath{\Lz^0_\bquark}}\xspace}

\def\Lc      {{\ensuremath{\Lz^+_\cquark}}\xspace}

\def\BF         {{\ensuremath{\mathcal{B}}}\xspace}

\def\BR         {\BF}
\newcommand{\decay}[2]{\ensuremath{#1\!\to #2}\xspace}         

\def\to                 {\ensuremath{\rightarrow}\xspace}

\def\Vud  {{\ensuremath{V_{\uquark\dquark}}}\xspace}

\def\Vus  {{\ensuremath{V_{\uquark\squark}}}\xspace}

\def\AT#1     {\ensuremath{A_{\mathrm{T}}^{#1}}\xspace}           

\def\C#1      {\ensuremath{\mathcal{C}_{#1}}\xspace}                       
\def\Cp#1     {\ensuremath{\mathcal{C}_{#1}^{'}}\xspace}                    
\def\Ceff#1   {\ensuremath{\mathcal{C}_{#1}^{\mathrm{(eff)}}}\xspace}        
\def\Cpeff#1  {\ensuremath{\mathcal{C}_{#1}^{'\mathrm{(eff)}}}\xspace}       
\def\Ope#1    {\ensuremath{\mathcal{O}_{#1}}\xspace}                       
\def\Opep#1   {\ensuremath{\mathcal{O}_{#1}^{'}}\xspace}                    

\newcommand{\tev}{\ifthenelse{\boolean{inbibliography}}{\ensuremath{~T\kern -0.05em eV}}{\ensuremath{\mathrm{\,Te\kern -0.1em V}}}\xspace}
\newcommand{\gev}{\ensuremath{\mathrm{\,Ge\kern -0.1em V}}\xspace}
\newcommand{\mev}{\ensuremath{\mathrm{\,Me\kern -0.1em V}}\xspace}
\newcommand{\kev}{\ensuremath{\mathrm{\,ke\kern -0.1em V}}\xspace}
\newcommand{\ev}{\ensuremath{\mathrm{\,e\kern -0.1em V}}\xspace}
\newcommand{\gevc}{\ensuremath{{\mathrm{\,Ge\kern -0.1em V\!/}c}}\xspace}
\newcommand{\mevc}{\ensuremath{{\mathrm{\,Me\kern -0.1em V\!/}c}}\xspace}
\newcommand{\gevcc}{\ensuremath{{\mathrm{\,Ge\kern -0.1em V\!/}c^2}}\xspace}
\newcommand{\gevgevcccc}{\ensuremath{{\mathrm{\,Ge\kern -0.1em V^2\!/}c^4}}\xspace}
\newcommand{\mevcc}{\ensuremath{{\mathrm{\,Me\kern -0.1em V\!/}c^2}}\xspace}

\def\mm   {\ensuremath{\mathrm{ \,mm}}\xspace}

\def\mum  {\ensuremath{{\,\upmu\mathrm{m}}}\xspace}

\def\invfb   {\ensuremath{\mbox{\,fb}^{-1}}\xspace}

\def\ps   {\ensuremath{{\mathrm{ \,ps}}}\xspace}

\newcommand{\chisq}{\ensuremath{\chi^2}\xspace}

\def\gsim{{~\raise.15em\hbox{$>$}\kern-.85em
          \lower.35em\hbox{$\sim$}~}\xspace}
\def\lsim{{~\raise.15em\hbox{$<$}\kern-.85em
          \lower.35em\hbox{$\sim$}~}\xspace}

\def\sPlot{\mbox{\em sPlot}\xspace}

\def\sqs   {\ensuremath{\protect\sqrt{s}}\xspace}

\def\ptot       {\mbox{$p$}\xspace}
\def\pt         {\mbox{$p_{\mathrm{ T}}$}\xspace}

\def\evtgen     {\mbox{\textsc{EvtGen}}\xspace}
\def\geant      {\mbox{\textsc{Geant4}}\xspace}
\def\pythia     {\mbox{\textsc{Pythia}}\xspace}

\def\tell1  {TELL1\xspace}
\def\ukl1   {UKL1\xspace}

\newcommand{\ie}{\mbox{\itshape i.e.}\xspace}

\newcommand{\Lcp}{\ensuremath{\Lc}\xspace}
\newcommand{\LzToppim}{\decay{\Lz}{\proton\pim}}
\newcommand{\DzPID}{\decay{\Dstarp}{\Dz \pip} (with \decay{\Dz}{\Km \pip})\xspace}
\newcommand{\LbToLcphh}{\decay{\Lb}{\Lcp (\proton h h^{'}) \mun \neumb}}
\newcommand{\LcpTopKmpip}{\decay{\Lcp}{\proton \Km \pip}}
\newcommand{\LcpToppimpip}{\decay{\Lcp}{\proton \pim \pip}}
\newcommand{\LcpTopKmKp}{\decay{\Lcp}{\proton \Km \Kp}}
\newcommand{\LcpToppimKp}{\decay{\Lcp}{\proton \pim \Kp}}
\newcommand{\LcpTophh}{\decay{\Lcp}{\proton \Ph \Ph'}}
\newcommand{\DpTopipKmpip}{\decay{\Dp}{\Km \pip \pip}}

\newcommand{\relBR}[2]{\ensuremath{\frac{\BR(#1)}{\BR(#2)}}\xspace}
\newcommand{\relBRinline}[2]{\ensuremath{{\BR(#1)}/{\BR(#2)}}\xspace}

\newcommand{\relBRKK}{\relBR{\LcpTopKmKp}{\LcpTopKmpip}}
\newcommand{\relBRpipi}{\relBR{\LcpToppimpip}{\LcpTopKmpip}}
\newcommand{\relBRDCS}{\relBR{\LcpToppimKp}{\LcpTopKmpip}}
\newcommand{\relBRKKinline}{\relBRinline{\LcpTopKmKp}{\LcpTopKmpip}}
\newcommand{\relBRpipiinline}{\relBRinline{\LcpToppimpip}{\LcpTopKmpip}}
\newcommand{\relBRDCSinline}{\relBRinline{\LcpToppimKp}{\LcpTopKmpip}}

\newcommand{\ip}{\ensuremath{\mathrm{IP}}\xspace}
\newcommand{\ipchisq}{\ensuremath{\chisq_{\ip}}\xspace}
\newcommand{\logipchisq}{\ensuremath{\ln\!\left(\ipchisq\right)}\xspace}

\newcommand{\Tabref}[1]{Table~\protect\ref{#1}}
\newcommand{\Figref}[1]{Figure~\protect\ref{#1}}
\newcommand{\Secref}[1]{Section~\protect\ref{#1}}

\newcommand{\rhoone}{\ensuremath{\Prho_{\mathrm{1}}}\xspace}
\newcommand{\rhotwo}{\ensuremath{\Prho_{\mathrm{2}}}\xspace}

\newcommand{\xone}{\ensuremath{\Px_{\mathrm{1}}}\xspace}
\newcommand{\xtwo}{\ensuremath{\Px_{\mathrm{2}}}\xspace}

\newcommand{\plamc}{\ensuremath{\mathrm{P}_{\Lc}}\xspace}

\newcommand{\mphone}{\ensuremath{m_{ph}}\xspace}
\newcommand{\mphtwo}{\ensuremath{m_{ph^{'}}}\xspace}
\newcommand{\mhh}{\ensuremath{m_{hh^{'}}}\xspace}

\usepackage{cite}
\usepackage{mciteplus}

\usepackage{blindtext}
\usepackage{enumitem}
\usepackage{longtable}

\begin{document}
\renewcommand{\thefootnote}{\fnsymbol{footnote}}
\setcounter{footnote}{1}

\newlength{\titlepageobjspc}
\setlength{\titlepageobjspc}{0.9cm}

\begin{titlepage}
\pagenumbering{roman}

\vspace*{-1.5cm}
\centerline{\large EUROPEAN ORGANIZATION FOR NUCLEAR RESEARCH (CERN)}
\vspace*{1.5cm}
\noindent
\begin{tabular*}{\linewidth}{lc@{\extracolsep{\fill}}r@{\extracolsep{0pt}}}
\ifthenelse{\boolean{pdflatex}}
{\vspace*{-2.7cm}\mbox{\!\!\!\includegraphics[width=.14\textwidth]{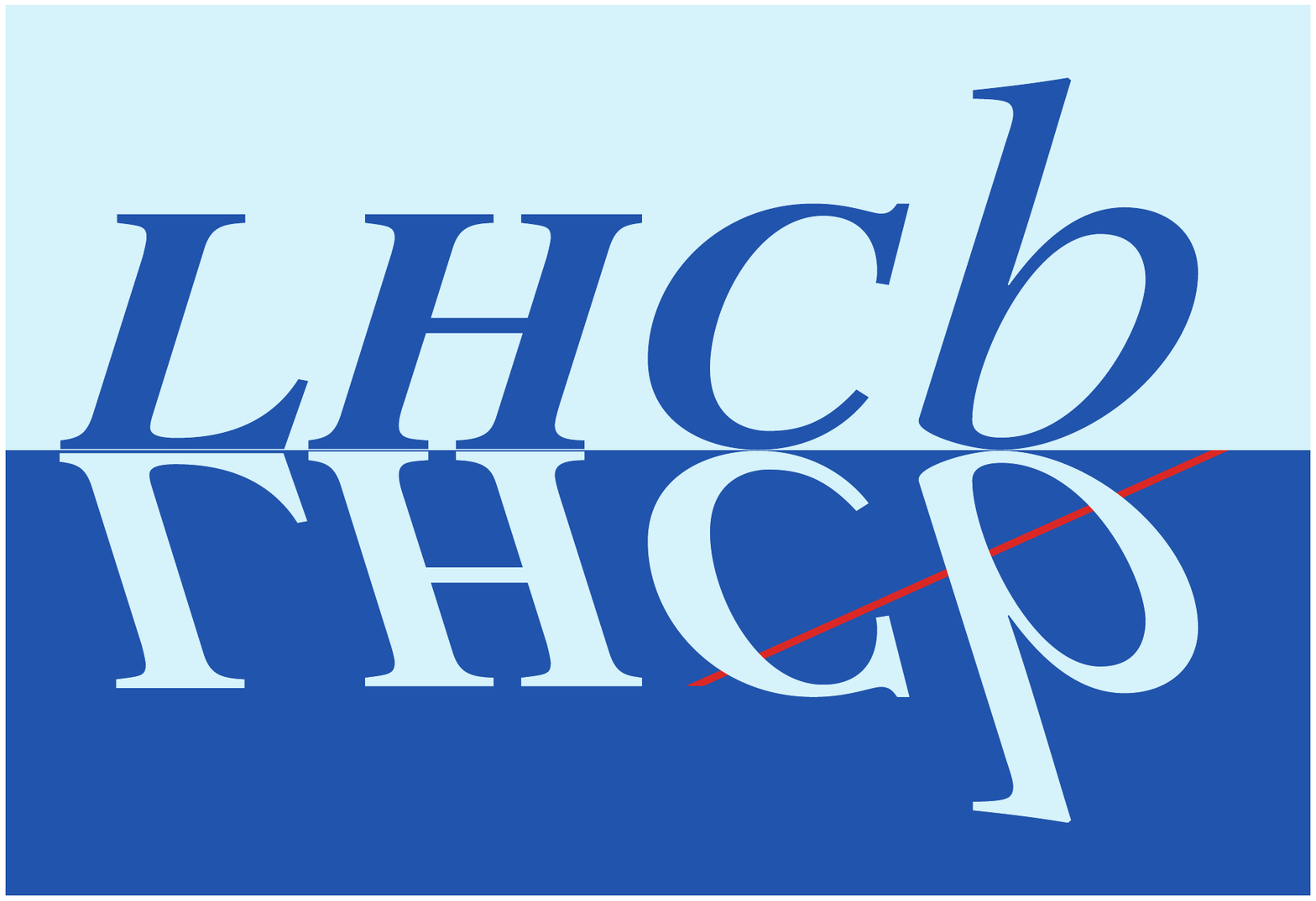}} & &}
{\vspace*{-1.2cm}\mbox{\!\!\!\includegraphics[width=.12\textwidth]{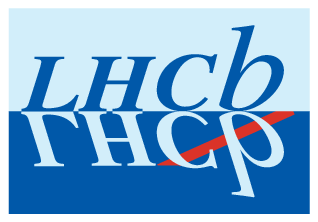}} & &}
\\
 & & CERN-EP-2017-247 \\  
 & & LHCb-PAPER-2017-026 \\  
 & & January 12, 2018 \\
 & & \\
\end{tabular*}

\vspace*{\titlepageobjspc}
\vspace{-8mm}
{\normalfont\bfseries\boldmath\huge
\begin{center}
  \papertitle 
\end{center}
}
\vspace*{\titlepageobjspc}
\begin{center}
\paperauthors\footnote{Authors are listed at the end of this paper.}
\end{center}

\vspace{\fill}
\begin{abstract}
  \noindent
  The ratios of the branching fractions of the decays \LcpToppimpip, \LcpTopKmKp, and \LcpToppimKp with respect to the Cabibbo-favoured \LcpTopKmpip decay are measured using proton-proton collision data collected with the \lhcb experiment at a $7\tev$ centre-of-mass energy and corresponding to an integrated luminosity of $1.0\invfb$:
  \begin{align*}
\relBRpipi             & = (7.44 \pm 0.08 \pm 0.18)\,\%, \\  	
\relBRKK              &= (1.70 \pm 0.03 \pm 0.03)\,\%, \\
\relBRDCS & = (0.165 \pm 0.015 \pm 0.005 )\,\%, 
\end{align*}
where the uncertainties are statistical and systematic, respectively.
These results are the most precise measurements of these quantities to date.
When multiplied by the world-average value for \mbox{$\mathcal{B}(\LcpTopKmpip)$}, the corresponding branching fractions are
\begin{align*}
\mathcal{B}(\LcpToppimpip) &= (4.72 \pm 0.05 \pm 0.11 \pm 0.25) \times 10^{-3}, \\
\mathcal{B}(\LcpTopKmKp) &= (1.08 \pm 0.02 \pm 0.02 \pm 0.06) \times 10^{-3}, \\
\mathcal{B}(\LcpToppimKp) &= (1.04 \pm 0.09 \pm 0.03 \pm 0.05) \times 10^{-4},
\end{align*}
where the final uncertainty is due to \mbox{$\mathcal{B}(\LcpTopKmpip)$}.
\end{abstract}
\vspace*{\titlepageobjspc}

\begin{center}
 Published in JHEP 03 (2018) 043
\end{center}

\vspace{\fill}
{\footnotesize 
\centerline{\copyright~\papercopyright, licence \href{\paperlicenceurl}{\paperlicence}.}}
\vspace*{2mm}

\end{titlepage}

\newpage
\setcounter{page}{2}
\mbox{~}
\cleardoublepage

\renewcommand{\thefootnote}{\arabic{footnote}}
\setcounter{footnote}{0}
\pagestyle{plain}
\setcounter{page}{1}
\pagenumbering{arabic}

\section{Introduction}
\label{sec:Introduction}
Nonleptonic decays of charmed baryons are a useful environment in which to study the interplay of the weak and strong interactions.
Measurements of their branching fractions are of great importance in understanding the internal dynamics of the decays.
The last few years have seen advances in the study of \LcpTophh decays, where
\mbox{$\Ph\Ph' \in \{\Km\pip, \Km\Kp, \pim\pip, \pim\Kp \}$}.
Until recently, measurements of the absolute branching fraction of the \LcpTopKmpip decay suffered from model dependence, relying on assumptions concerning several \Bbar, \Lc and \Dp branching fraction ratios and decay widths.
The first model-independent measurements of the absolute branching fraction of the \LcpTopKmpip decay have been made by the Belle~\cite{Zupanc:2013iki} and BESIII~\cite{Ablikim:2015flg} collaborations.
The precision of a number of \Lcp decay branching fractions has also been improved at the \B factories~\cite{Abe:2001mb, Ablikim:2015flg,Ablikim:2015prg, Ablikim:2016tze}, while the first measurement of a doubly Cabibbo-suppressed (DCS) charmed-baryon decay, \LcpToppimKp, has been performed by the \belle collaboration~\cite{Yang:2015ytm}.

Unlike in the charmed-meson sector, there exist a large number of favoured internal  \W-boson exchange decays which can be readily studied.
Quark-level diagrams demonstrating external \W-emission, internal \W-emission, and \W-exchange are shown in \Figref{fig:quarks}.
\begin{figure}[htb]
	\centering
	\scriptsize
	\subfloat[ ]{\begin{tikzpicture}
  \begin{feynman}[small]
  
	\vertex(c1){\cquark};
	\vertex[right=2cm of c1](c2) ;
	\vertex[right=2cm of c2](c3){\squark};

	\vertex[above right=1cm of c2](w1);
	\vertex[above=2.0em of c3](w2){\dquarkbar};
	\vertex[above=1.5em of w2](w3){\uquark};

	\vertex[below=0.5cm of c3](q1){\uquarkbar};
	\vertex[below=2.1em of q1](q2){\uquark};

    \vertex[below=1em of c1] (a1) {\uquark};
	\vertex[below=1em of q2] (a2){\uquark};
	\vertex[below=2em of c1] (b1) {\dquark};
	\vertex[below=1em of a2] (b2) {\dquark};
	
	\vertex[above left=0.5cm of q2] (qb);
	
   \diagram* {
      {[edges=fermion]
        (a1) -- (a2),
        (b1) -- (b2),
        (c1) -- (c2) -- (c3),
        (w1) -- (w3),
         (w2) -- (w1);
        (q1) -- [half right] (q2)
	};
         (c2) -- [boson, edge label=\Wp](w1) ;
  };

    \draw [decoration={brace}, decorate]  (q2.north east) -- (b2.south east)
          node [pos=0.5, right] {\ \proton};
          
     \draw [decoration={brace}, decorate]  (w3.north east) -- (w2.south east)
          node [pos=0.5, right] {\ \pip};

     \draw [decoration={brace}, decorate]  (c3.north east) -- (q1.south east)
		node [pos=0.5, right] {\ \Km};

    \draw [decoration={brace}, decorate]  (b1.south west) -- (c1.north west)
node [pos=0.5, left] {\Lcp \ };          
          
  \end{feynman}
\end{tikzpicture}}
	\subfloat[ ]{\begin{tikzpicture}
  \begin{feynman}[small]
  
	\vertex(c1){\cquark};
	\vertex[right=2cm of c1](c2) ;
	\vertex[right=2cm of c2](c3){\dquark};

	\vertex[above right=1cm of c2](w1);
	\vertex[above=2.0em of c3](w2){\squarkbar};
	\vertex[above=1.5em of w2](w3){\uquark};

	\vertex[below=0.5cm of c3](q1){\uquarkbar};
	\vertex[below=2.1em of q1](q2){\uquark};

    \vertex[below=1em of c1] (a1) {\uquark};
	\vertex[below=1em of q2] (a2){\uquark};
	\vertex[below=2em of c1] (b1) {\dquark};
	\vertex[below=1em of a2] (b2) {\dquark};
	
	\vertex[above left=0.5cm of q2] (qb);
	
   \diagram* {
      {[edges=fermion]
        (a1) -- (a2),
        (b1) -- (b2),
        (c1) -- (c2) -- (c3),
        (w2) -- (w1),
        (w1) -- (w3),
        (q1) -- [half right] (q2)
	};
         (c2) -- [boson, edge label=\Wp](w1) ;
  };

    \draw [decoration={brace}, decorate]  (q2.north east) -- (b2.south east)
          node [pos=0.5, right] {\ \proton};
          
     \draw [decoration={brace}, decorate]  (w3.north east) -- (w2.south east)
          node [pos=0.5, right] {\ \Kp};

     \draw [decoration={brace}, decorate]  (c3.north east) -- (q1.south east)
		node [pos=0.5, right] {\ \pim};

    \draw [decoration={brace}, decorate]  (b1.south west) -- (c1.north west)
node [pos=0.5, left] {\Lcp \ };          
          
  \end{feynman}
\end{tikzpicture}} \\ 
	\subfloat[ ]{\begin{tikzpicture}
  \begin{feynman}
  
	\vertex(c1){\cquark};
	\vertex[right=2cm of c1](c2) ;
	\vertex[right=2cm of c2](c3){\uquark};

	\vertex[below right=1cm of c2](w1);

	\vertex[above=2.0em of c3](w2){\uquarkbar};
	\vertex[above=1.5em of w2](w3){\squark};

	\vertex[below=0.5cm of c3](q1){\dquarkbar};
	\vertex[below=2.1em of q1](q2){\uquark};

    \vertex[below=1em of c1] (a1) {\uquark};
	\vertex[below=1em of q2] (a2){\uquark};
	\vertex[below=2em of c1] (b1) {\dquark};
	\vertex[below=1em of a2] (b2) {\dquark};
	
	\vertex[above left=0.5cm of q2] (qb);

	\vertex[below  left=1.8cm of w3](we);
	
   \diagram* {
      {[edges=fermion]
        (a1) -- (a2),
        (b1) -- (b2),
        (c1) -- (c2) -- (w3)
        (q1) -- (we)
        (we) -- (q2)
        (w2) -- [half right] (c3)
	};
      (c2) -- [boson, edge label=\Wp](we) ;
  };

    \draw [decoration={brace}, decorate]  (q2.north east) -- (b2.south east)
          node [pos=0.5, right] {\ \proton};
          
     \draw [decoration={brace}, decorate]  (w3.north east) -- (w2.south east)
          node [pos=0.5, right] {\ \Km};

     \draw [decoration={brace}, decorate]  (c3.north east) -- (q1.south east)
		node [pos=0.5, right] {\ \pip};

    \draw [decoration={brace}, decorate]  (b1.south west) -- (c1.north west)
node [pos=0.5, left] {\Lcp \ };          
          
  \end{feynman}
\end{tikzpicture}}
	\subfloat[ ]{\begin{tikzpicture}
  \begin{feynman}
  
	\vertex(c1){\cquark};
	\vertex[right=2cm of c1](c2) ;
	\vertex[right=2cm of c2](c3){\uquark};

	\vertex[below right=1cm of c2](w1);

	\vertex[above=2.0em of c3](w2){\uquarkbar};
	\vertex[above=1.5em of w2](w3){\dquark};

	\vertex[below=0.5cm of c3](q1){\squarkbar};
	\vertex[below=2.1em of q1](q2){\uquark};

    \vertex[below=1em of c1] (a1) {\uquark};
	\vertex[below=1em of q2] (a2){\uquark};
	\vertex[below=2em of c1] (b1) {\dquark};
	\vertex[below=1em of a2] (b2) {\dquark};
	
	\vertex[above left=0.5cm of q2] (qb);

	\vertex[below  left=1.8cm of w3](we);
	
   \diagram* {
      {[edges=fermion]
        (a1) -- (a2),
        (b1) -- (b2),
        (c1) -- (c2) -- (w3)
        (q1) -- (we)
        (we) -- (q2)
        (w2) -- [half right] (c3)
	};
      (c2) -- [boson, edge label=\Wp](we) ;
  };

    \draw [decoration={brace}, decorate]  (q2.north east) -- (b2.south east)
          node [pos=0.5, right] {\ \proton};
          
     \draw [decoration={brace}, decorate]  (w3.north east) -- (w2.south east)
          node [pos=0.5, right] {\ \pim};

     \draw [decoration={brace}, decorate]  (c3.north east) -- (q1.south east)
		node [pos=0.5, right] {\ \Kp};

    \draw [decoration={brace}, decorate]  (b1.south west) -- (c1.north west)
node [pos=0.5, left] {\Lcp \ };          
          
  \end{feynman}
\end{tikzpicture}} \\ 
	\subfloat[ ]{\begin{tikzpicture}
  \begin{feynman}
  
	\vertex(c1){\cquark};
	\vertex[right=2cm of c1](c2) ;
	\vertex[right=2cm of c2](c3){\uquark};

	\vertex[below right=1cm of c2](w1);

	\vertex[above=2.0em of c3](w2){\uquarkbar};
	\vertex[above=1.5em of w2](w3){\squark};

	\vertex[below=0.5cm of c3](q1){\dquarkbar};
	\vertex[below=2.1em of q1](q2){\dquark};

    \vertex[below=1em of c1] (a1) {\dquark};
	\vertex[below=1em of q2] (a2){\uquark};
	\vertex[below=2em of c1] (b1) {\uquark};
	\vertex[below=1em of a2] (b2) {\uquark};
	
	\vertex[above left=0.5cm of q2] (qb);

	\vertex[below  left=1.8cm of w3](we);
	\vertex[below =1.0em of c2](we2);
	
   \diagram* {
      {[edges=fermion]
        (a1) -- (we2) -- (a2),
        (b1) -- (b2),
        (c1) -- (c2) -- (w3)
        (w2) -- [half right] (c3)  
                (q1) -- [half right] (q2)
	};
      (c2) -- [boson, edge label=\Wp](we2) ;
  };

    \draw [decoration={brace}, decorate]  (q2.north east) -- (b2.south east)
          node [pos=0.5, right] {\ \proton};
          
     \draw [decoration={brace}, decorate]  (w3.north east) -- (w2.south east)
          node [pos=0.5, right] {\ \Km};

     \draw [decoration={brace}, decorate]  (c3.north east) -- (q1.south east)
		node [pos=0.5, right] {\ \pip};

    \draw [decoration={brace}, decorate]  (b1.south west) -- (c1.north west)
node [pos=0.5, left] {\Lcp \ };          
          
  \end{feynman}
\end{tikzpicture}}
	\caption[]{Weak decays of \Lcp to a proton and two mesons, without hyperon mediation. Shown are external \W-emission for (a) \LcpTopKmpip and (b) \LcpToppimKp, internal \W-emission for (c) \LcpTopKmpip and (d) \LcpToppimKp, and \W-exchange for (e) \LcpTopKmpip.
		\label{fig:quarks}}
\end{figure}
As can be seen, while \W-boson exchange is not permitted in the decay \LcpToppimKp, it is allowed in the decay \LcpTopKmpip.
The ratio of the branching fractions \relBRDCSinline is a useful variable with which to  indirectly study the role of \W-boson exchange in hadronic decays.
In the absence of flavour-SU(3) symmetry breaking, the ratio can naively be expected to be equal to $\tan^{4}\theta_{c}$~\cite{Lipkin:2002za},  where $\theta_c$ is the Cabibbo mixing angle~\cite{Cabibbo:1963yz}.
Taking the most recent measurements of $|\Vud|$ and $|\Vus|$~\cite{PDG2016} yields a value $\tan^{4}\theta_{c} \approx 0.285\,\%$.
The Belle measurement for \relBRDCSinline corresponds to $(0.82 \pm 0.12) \tan^{4}\theta_{c}$.

In this paper we report measurements of the ratios of the branching fractions
\begin{align*}
\relBRKK, \
\relBRpipi \ \mathrm{and} \
\relBRDCS.
\end{align*}
These measurements are carried out using a data sample, corresponding to an integrated luminosity of 1.0~\invfb of \proton\proton collision data, collected with the \lhcb detector at a centre-of-mass energy of $\sqs=7\tev$.
The \Lcp candidates are reconstructed in semileptonic (SL) decays of $\decay{\Lb}{\Lc \mun X}$, where $X$ is any particle in this decay that is not reconstructed.
These decays have a low level of background due to the use of high-purity muon triggers and the displacement of the \Lcp production point from the primary \proton\proton collision.
As a powerful cross-check, the same measurements, although with a lower precision, are carried out using a sample of \Lcp produced in the primary \proton\proton interaction vertex (PV), referred to as the prompt sample.

\section{Detector and simulation}
\label{sec:Detector}

The \lhcb detector~\cite{LHCb-DP-2014-002} is a single-arm forward
spectrometer covering the \mbox{pseudorapidity} range $2<\eta <5$,
designed for the study of particles containing \bquark or \cquark
quarks. The detector includes a high-precision tracking system
consisting of a silicon-strip vertex detector surrounding the $pp$
interaction region, a large-area silicon-strip detector located
upstream of a dipole magnet with a bending power of about
$4{\mathrm{\,Tm}}$, and three stations of silicon-strip detectors and straw
drift tubes placed downstream of the magnet.
The tracking system provides a measurement of momentum, \ptot, of charged particles with
a relative uncertainty that varies from 0.5\,\% at low momentum to 1.0\,\% at 200\gevc.
The minimum distance of a track to a primary vertex, the impact parameter (IP), 
is measured with a resolution of $(15+29/\pt)\mum$,
where \pt is the component of the momentum transverse to the beam, in\,\gevc.
Different types of charged hadrons are distinguished using information
from two ring-imaging Cherenkov (RICH) detectors~\cite{LHCb-DP-2012-003}, allowing for an effective discrimination between the different \LcpTophh final states.
Photons, electrons and hadrons are identified by a calorimeter system consisting of
scintillating-pad and preshower detectors, an electromagnetic
calorimeter and a hadronic calorimeter.
Muons are identified by a
system composed of alternating layers of iron and multiwire
proportional chambers.

The online event selection is performed by a trigger, which  consists of  a  hardware stage, based  on information from the calorimeter and muon systems, followed by a software stage which is divided into two parts.
The first employs a partial reconstruction of the candidates from the hardware trigger and a cut-based selection, while the second utilises a full event reconstruction and further, often more complex, selection criteria on candidates.
Selection requirements can be made on whether a trigger decision was satisfied by any given object in the event (including nonsignal objects).
In the offline selection, trigger decisions are associated with reconstructed particles.
Therefore requirements can be made on whether the signal candidate was responsible for satisfying the trigger decision, or if another nonsignal object in the event satisfied the trigger decision, or a combination of the two.
The detailed trigger requirements for the semileptonic and prompt samples are described in \Secref{sec:Selection}.

In the simulation, $pp$ collisions are generated using
\pythia~\cite{Sjostrand:2006za,*Sjostrand:2007gs} 
with a specific \lhcb
configuration~\cite{LHCb-PROC-2010-056}.
The heavy flavour decays
are described by \evtgen~\cite{Lange:2001uf} with the decay kinematics of the \LcpTophh
generated according to a phase-space distribution.
The interaction of the generated particles with the detector, and its response,
are implemented using the \geant toolkit~\cite{Allison:2006ve} as described in
Ref.~\cite{LHCb-PROC-2010-056}.

\section{Candidate selection}
\label{sec:Selection}
The different production mechanisms in the SL and prompt processes necessitate two distinct selections, which are verified to result in statistically independent samples of \Lcp candidates.
The selections are developed using a fraction of the \LcpTopKmpip data corresponding to 10\,\% of the integrated luminosity, chosen randomly.
This sample is then discarded from measurements of the ratios of branching fractions, with an appropriate scaling factor applied to the final results.

\subsection{{\protect\boldmath\LbToLcphh} selection}
The trigger selection at the hardware stage and the first software stage is focussed upon the muon in the \Lb decay, such that the dependence of the selection upon the \Lcp decay product kinematics is reduced.
This results in the ratios of trigger acceptance efficiencies between the \LcpTophh modes being uniform at these stages of the trigger.
The muon candidate is required to have a $\pt > 1.7\gevc$ and to be responsible for the decision of both the hardware trigger and the first stage of the software trigger.
The latter uses additional detector information to confirm that the muon has a high \pt and is significantly displaced from the primary vertex.
In the second stage of the software trigger, a general algorithm designed for identifying semileptonic \bquark-hadron decays selects \LbToLcphh candidates, requiring a high \pt muon that is significantly displaced from the PV.
This muon must then form a displaced secondary vertex with between one and three other tracks.
This vertex must have at least one track with $\pt > 1.7\gevc$ and \ipchisq with respect to any PV greater than 16, where \ipchisq is defined as the difference in the fit \chisq of a given PV reconstructed with and without the considered particle.

The candidates selected by the trigger are then filtered to improve the signal purity.
Charged hadrons are selected with a momentum  $\ptot > 2.0\gevc$, and $\pt>0.3\gevc$.
All tracks must have $\ipchisq > 9$ such that they are significantly displaced from any PV in the event, and have a good fit quality.
Three such tracks must then form a high-quality vertex with a flight-distance-significance greater than 100 (defined as the measured flight distance from any PV divided by its uncertainty).
The \pt of the three-particle combination must also be greater than $1.8\gevc$.

Particle identification (PID)
is applied to each charged hadron in order to select exclusive samples of each final state, and to reject backgrounds from other multibody charm decays.
Tight PID selection criteria are enforced on the proton and kaon candidates in order to suppress possible backgrounds from misidentified \cquark-hadron decays, with a weaker requirement placed upon the pion candidates.

Muon candidates must have a high-quality track fit, and have $\ipchisq>9$, $\ptot > 3\gevc$ and $\pt > 0.8\gevc$.
A moderate PID requirement
is also enforced to reduce the background from $\pi-\mu$ misidentification.
Finally, the muon and \Lcp candidates are required to form a common vertex with a fit \chisq lower than 6.
The invariant mass of the three tracks in the \Lcp combination is required to be within $\pm 40$\mevcc of the known \Lcp mass~\cite{PDG2016}.
The invariant mass of the combination of the muon and the \Lcp candidate must fall in the range $2.5 - 6.0\gevcc$.

\subsection{Prompt {\protect\boldmath\LcpTophh} selection}
To ensure that the trigger acceptance does not depend upon the \Lcp decay channel and kinematics, all accepted candidates must have been triggered independently of the \Lcp decay products.
If the measured branching fraction ratios between the prompt and SL analyses are compatible, this is a strong indication of the robustness of our method given the very different triggering strategies.

To improve the sample purity a selection using PID and kinematic information is employed.
All charged hadrons forming the \Lcp candidate must have a momentum greater than $5\gevc$ and a \pt greater than $0.4\gevc$, and at least one hadron is required to have a \pt exceeding $1.2\gevc$.
All hadronic tracks should have a \ipchisq greater than 4, with at least one greater than 8.
All tracks should have a good fit quality.
The PID requirements on the protons, kaons and pions in the selection are identical to those used in the SL analysis.

The \Lcp candidate formed from these tracks is required to have a vertex-fit \chisq lower than 20, and a maximum distance-of-closest-approach between any two pairings of the decay products of $0.1\mm$.
The flight-distance significance is required to be greater than 16.
The reconstructed proper time of the \Lcp is required to be below $1.2\ps$ to reject misreconstructed charmed-meson decays and \Lcp produced in decays of \bquark hadrons (referred to as secondary \Lcp).
The invariant mass of the \Lcp candidate is required to be within $\pm40\mevcc$ of the known \Lcp mass~\cite{PDG2016}.
Finally, the angle between the line joining the production and decay vertices and the reconstructed \Lcp momentum vector must be small.

\subsection{Selection efficiencies}
The efficiencies for the selection of signal decays are factorised into components which can be measured independently.
These efficiencies are the probability for the decays to occur within the detector acceptance, for the trigger to accept the signal event, for the final-state particles to be reconstructed, and for the decay to be selected.

The efficiency components are evaluated using simulation, with the exception of the PID selection efficiency, where a data-driven approach is utilised.
High-purity calibration samples of kaons and pions are acquired using \DzPID decays, which are identified without the use of PID requirements, while corresponding samples of protons are acquired using \LzToppim decays.
In the prompt analysis a small supplementary sample of \LcpTopKmpip decays is also used to acquire calibration protons, which are verified to be statistically independent of the signal \LcpTopKmpip due largely to their different triggering and selection strategies.
These calibration samples allow for an evaluation of the PID performance as a function of a set of variables which can fully characterise the single-track PID performance; in this analysis the track momentum and pseudorapidity are utilised.
The distributions of these variables for the \LcpTophh decay product tracks are then extracted using the \sPlot technique~\cite{Pivk:2004ty}, with the \Lcp candidate invariant mass as a discriminating variable.
A weighting procedure is then used to align the signal and calibration samples such that an average PID selection efficiency for the decay mode can be determined entirely through the use of data.

For the efficiencies determined from simulation, it is necessary to
consider the
unknown resonant structure of the \LcpTophh decays.
It is assumed that the decay is characterised both by intermediate two-body resonances and a nonresonant decay amplitude which is constant across the phase space.
According to the helicity formalism detailed in Ref.~\cite{Aitala:1999uq}, the differential decay rate as a function of the \Lc polarisation, \plamc, can be expressed as
\begin{align*}
\begin{split}
\rm{d}\Gamma & \sim \frac{1+\plamc}{2} \left( \left| \sum_{r}^{} BW (m_r) \alpha_{r, \frac{1}{2}, \frac{1}{2}}\right|^2 + \left|\sum_{r}^{} BW (m_r) \alpha_{r, \frac{1}{2}, -\frac{1}{2}}\right|^2 \right) \\
& + \frac{1-\plamc}{2} \left( \left| \sum_{r}^{} BW (m_r) \alpha_{r, -\frac{1}{2}, \frac{1}{2}}\right|^2 + \left|\sum_{r}^{} BW (m_r) \alpha_{r, -\frac{1}{2}, -\frac{1}{2}}\right|^2 \right)
\end{split}
\end{align*}
where $\Palpha_{r, \Pm, \lambda_{\rm{p}}}$ is the complex decay amplitude for resonance \Pr with spin \Pm (the \Lc spin projection onto the \Pz-axis), $\lambda_{\rm{p}}$ is the proton helicity  in the rest frame of the \Lc, and
$BW$ is the Breit-Wigner amplitude (the form of which may be found in Ref.~\cite{Frabetti:1994di}).
The \Lcp polarisation has not yet been measured at the \lhc.
For the prompt candidates, the polarisation axis is defined as the cross product of the beam momentum and the \Lcp momentum in the lab frame.
For the SL candidates, it is defined as the cross product of the \Lb momentum and the \Lc momentum in the lab frame.
The minimum parameterisation of this differential decay rate is represented by five kinematic variables.
These are any two of the following three invariant mass variables and each of the three angular variables, where each angle is defined in the \Lc rest frame:
\begin{description}[align=right,labelwidth=1.25cm]
	\item[\mphone] - the two-body invariant mass of the proton and the opposite-sign meson. \item [\mphtwo] - the two-body invariant mass of the proton and the same-sign meson.
	\item [\mhh] - the two-body invariant mass of the meson pair.
	\item[$\theta_{p}$] - the angle between the proton momentum vector and the polarisation axis of the \Lc.
	\item[$\phi_{p}$] - the angle between the 
	component of the proton momentum perpendicular to the \Lc polarisation axis and the direction of the \Lc momentum vector in the laboratory frame.
	\item[$\phi_{h_{1}h_{2}}$] - the angle between the plane containing the proton momentum vector and the \Lc polarisation vector, and the plane containing the two meson momentum vectors.
\end{description}

Some of the factorisable components of the selection efficiency depend upon combinations of these variables.
For each such dependence, the variable distributions from those listed above are obtained from the data using the \sPlot technique.
The simulated data is binned in these variables, and local efficiencies across the phase space are determined and applied to data on a per-candidate basis.
This procedure accurately describes the selection efficiencies using the simulated data without \emph{a priori} knowledge of the resonant structure of the \LcpTophh decays.
	For all schemas it is ensured that the signal yield in each bin, as determined with the \sPlot technique, is greater than three times its statistical error.
	Due to the finite size of the simulated sample, two-dimensional binnings are used in the final results, where the variables are chosen to be those with the greatest disagreement between data and simulation, as determined with a \chisq compatibility test.
	As a cross-check, three-dimensional binning schemas are implemented using these two primary variables in conjunction with every possible third variable, and in all cases the reweighted efficiencies are observed to be compatible with the two-dimensional binnings.

The full selection efficiency ratios for each measurement are summarised in \Tabref{tab:effratios}.
	As the number of kaons in the final state increases, the momentum available to the final state particles decreases, and the selection removes a higher fraction of the signal.
	The efficiency ratios are further from unity for the prompt measurements than for the corresponding SL measurements due to the tighter kinematic requirements used in the selection of the \LcpTophh decay products.
	The selection efficiencies display the same hierarchy before and after the reweighting procedure.

\begin{table}[htb]
	\centering
		\caption[Selection efficiency ratios]{%
		\small Selection efficiency ratios in the prompt and SL measurements, with their associated systematic uncertainties.
		$\epsilon_{\mathrm{CF}} / \epsilon_{\mathrm{CS}} $ denotes the ratio of the Cabibbo-favoured selection efficiency over that of the Cabibbo-suppressed mode.
		\label{tab:effratios}}%
	\begin{tabular}{l|l|r}
& Measurement 		& $\epsilon_{\mathrm{CF}} / \epsilon_{\mathrm{CS}} $ \\ \cline{2-3}
\multirow{2}{*}{Prompt}  & \relBRpipiinline	 & $0.67 \pm 0.02$ \\ 
& \relBRKKinline 		& $1.42 \pm 0.05$ \\ \hline
\multirow{4}{*}{SL} 
& \relBRpipiinline	 	& $0.96 \pm 0.02$ \\
& \relBRKKinline 		& $1.25 \pm 0.02$ \\
& \relBRDCSinline 	& $1.06 \pm 0.03$ \\
\end{tabular}

\end{table}

\section{Signal yield determination}
\label{sec:yield}
In both the SL and prompt analyses no contamination from backgrounds due to
misidentified charm decays, such as \DpTopipKmpip, is found in the data.
Cross-feed between the \LcpTophh modes, along with any contamination in the \LcpToppimKp or \LcpToppimpip channels from hyperon or \KS mediation, is also found to be negligible.
It is determined that the only decays left in the retained \Lc candidates are genuine \LcpTophh decays and backgrounds from combinations of unrelated tracks.

\subsection{{\protect\boldmath\LbToLcphh} yield determination}
\begin{figure}[htb]
	\captionsetup[subfigure]{labelformat=empty}	
	\centering
	\subfloat[  ]{\label{fig:slyields:z1}{\includegraphics[width=0.48\textwidth]{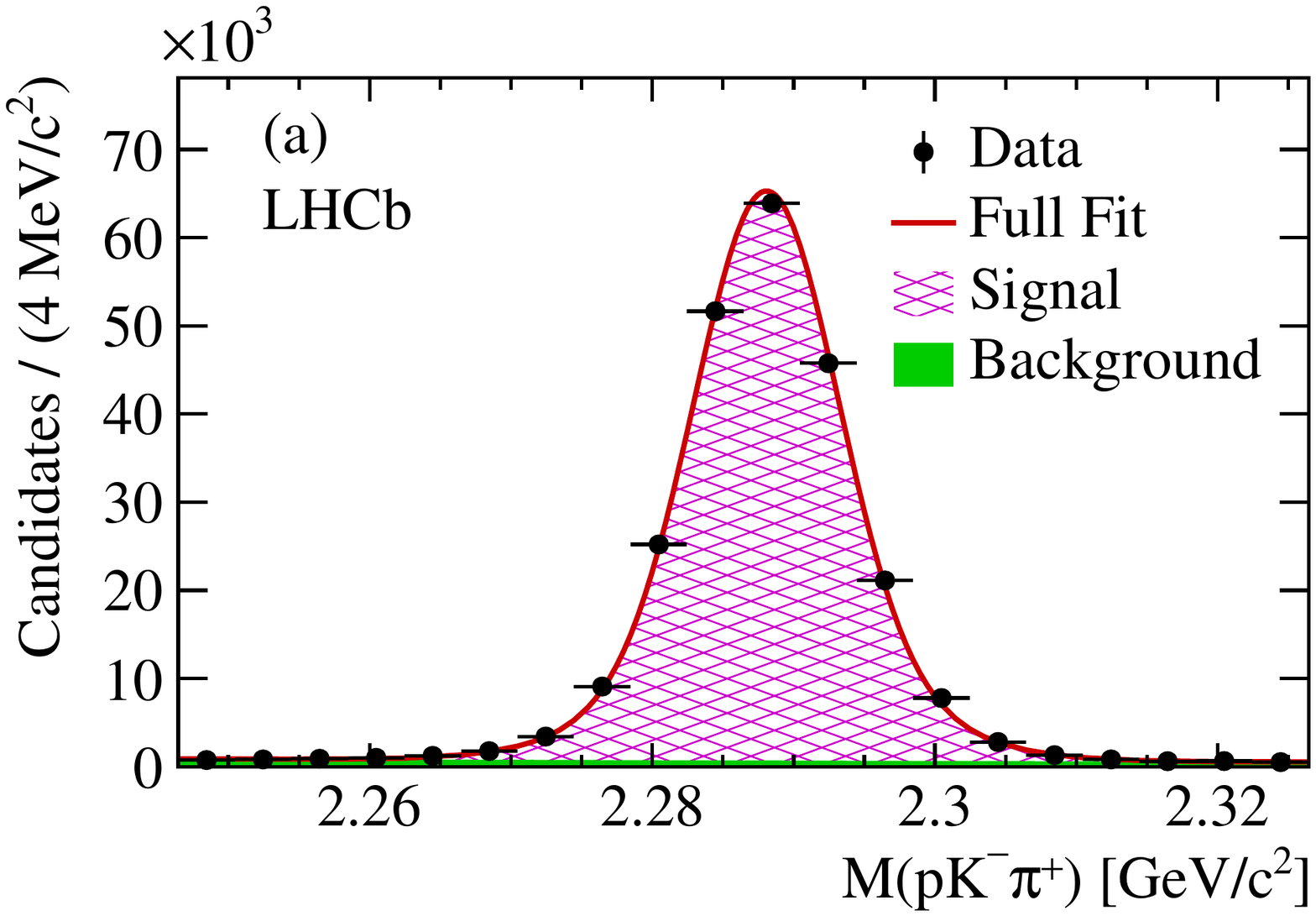}}}
	\subfloat[  ]{\label{fig:slyields:z2}{\includegraphics[width=0.48\textwidth]{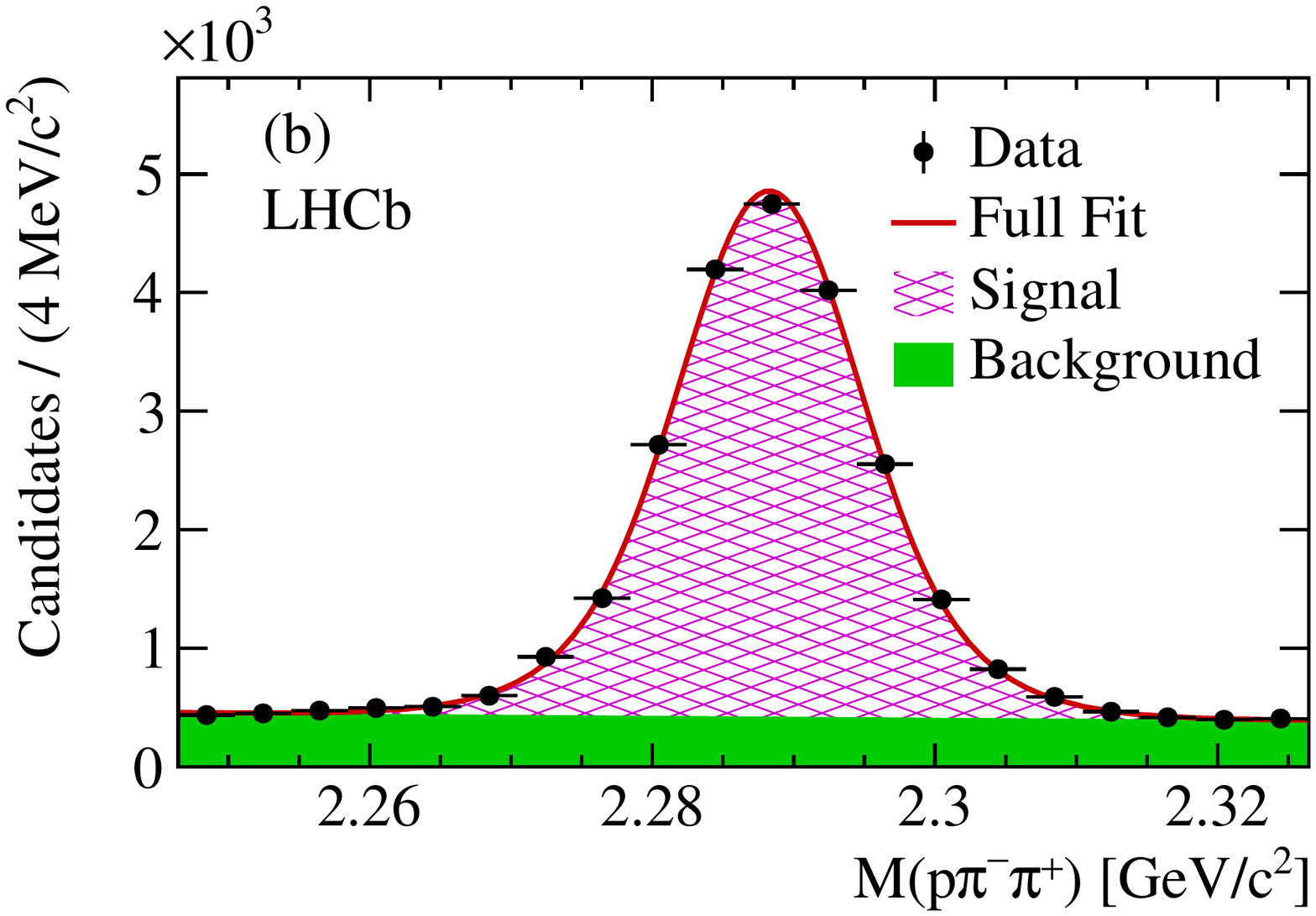}}} \\
	\vspace{-0.8cm}
	\subfloat[  ]{\label{fig:z3}{\includegraphics[width=0.48\textwidth]{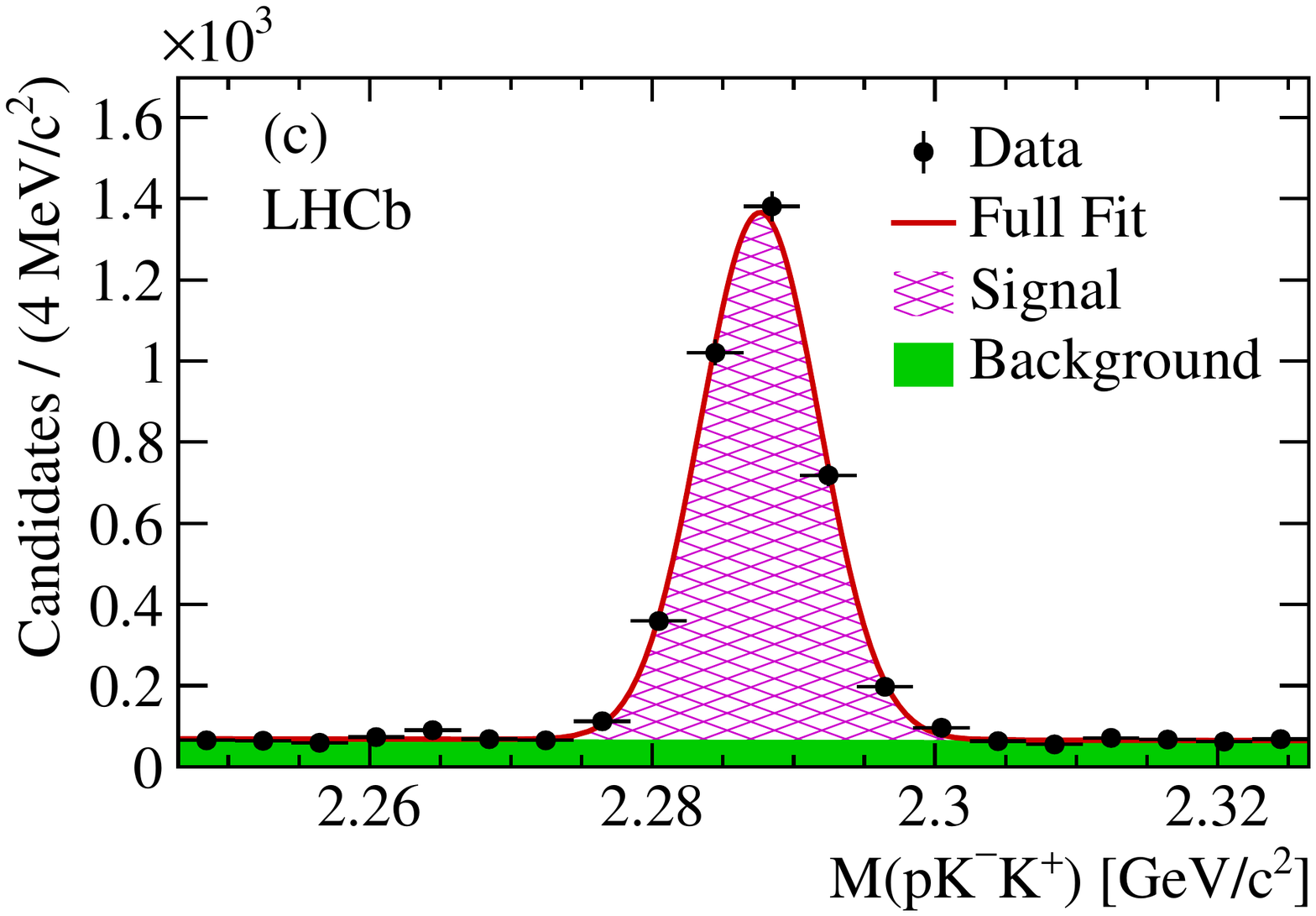}}}
	\subfloat[  ]{\label{fig:z4}{\includegraphics[width=0.48\textwidth]{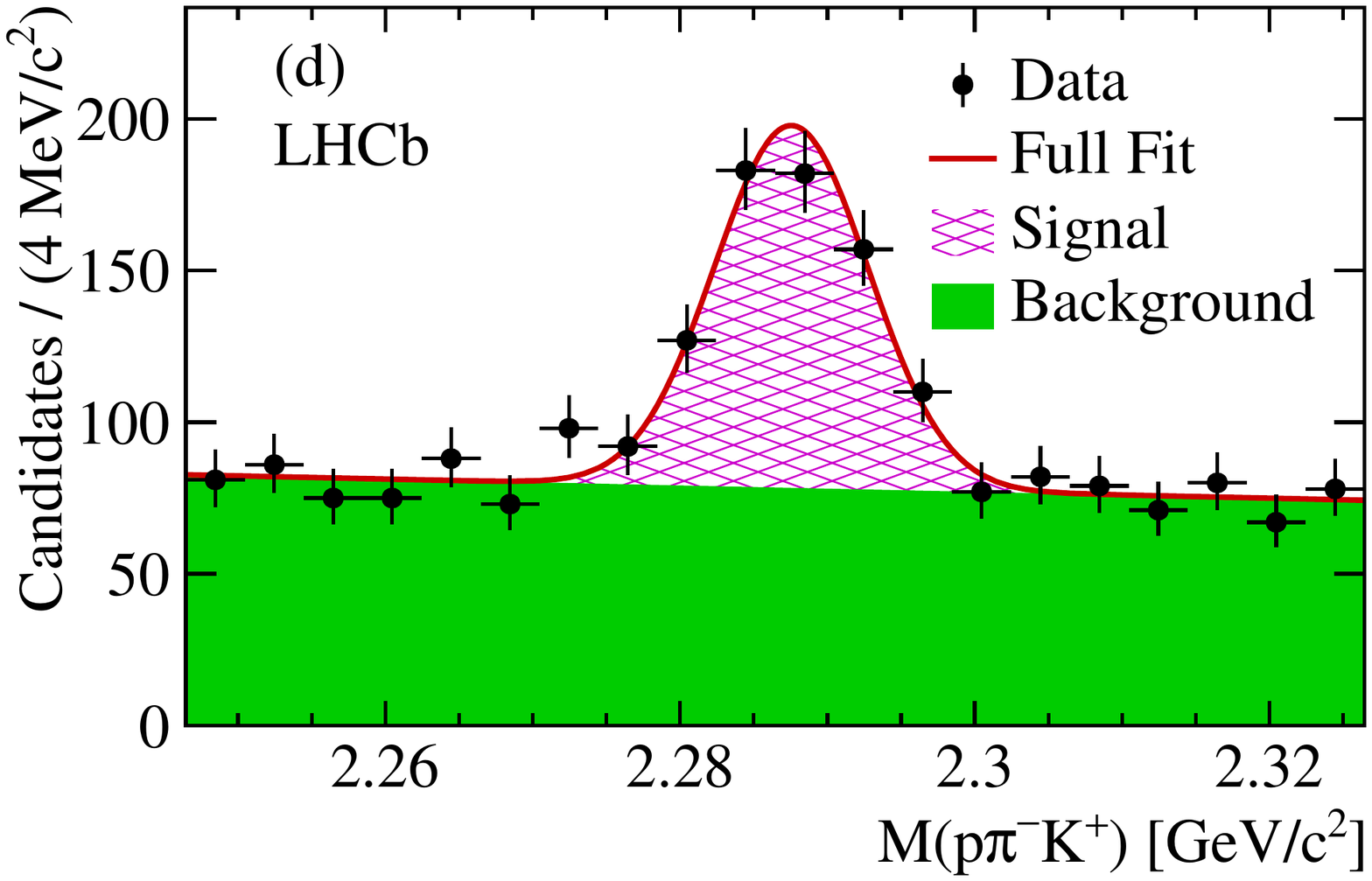}}} 
	\caption[]{
		Invariant mass distributions of (a) \LcpTopKmpip, (b) \LcpToppimpip, (c) \LcpTopKmKp, and (d) \LcpToppimKp decays, with fit results superimposed. The hatched magenta region indicates the signal, the shaded green region indicates the background from unrelated tracks, and the solid red line indicates the full fit.
		\label{fig:slyields}}
\end{figure}

The yields of each decay mode are extracted through an extended unbinned maximum likelihood fit to the \Lc invariant mass distributions.
The signal model for the \mbox{\LcpTopKmKp} and \LcpToppimKp modes is a Gaussian function, while for the \LcpTopKmpip and \LcpToppimpip modes the sum of two Gaussian functions with a common mean is used to account for the dependence of the reconstructed invariant mass resolution on the track momenta, which degrades the fit quality for a single Gaussian function in high-yield channels.
In all modes, the background model is an exponential function.
All parameters are free to vary in the fit.
The invariant mass distributions for each of the \LcpTophh modes, with the fit results overlaid, are shown in \Figref{fig:slyields}, and the signal yields are given in \Tabref{tab:yields}.

\subsection{Prompt {\protect\boldmath\LcpTophh} yield determination}
The yield determination procedure in the case of the prompt \Lcp is complicated by the presence of a large secondary \Lcp contribution.
These secondary \Lcp are statistically independent of the \Lcp selected in the SL analysis due to the different triggering and selection techniques employed.
The secondary \Lcp have different kinematic distributions than the prompt \Lcp.
Due to the kinematic criteria employed in the selection, the efficiency ratios between the \LcpTophh modes therefore vary between prompt and secondary \Lcp, resulting in the need to disentangle the prompt and secondary \Lcp candidates
Such a separation is achieved through examination of the \ipchisq of the \Lcp candidates.
The inclusion of a truly prompt \Lcp in the PV reconstruction generally results in a smaller increase of the PV-fit \chisq than in the case of an inclusion of a truly secondary \Lcp candidate.
To separate prompt and secondary \Lcp candidates
the natural logarithm of this quantity, \logipchisq, is utilised.

The yield determination in this case follows a two-step procedure.
First, the total number of \Lcp of each decay mode, \ie the sum of prompt and secondary \Lcp, is evaluated through an extended unbinned maximum likelihood fit to the \Lc invariant mass distributions.
This allows the \Lcp to be well separated from the combinatoric background.
The models used to describe the signal and background components are the same as for the \LbToLcphh analysis.
An unbinned extended maximum likelihood fit to the \Lc \logipchisq distributions is then performed, which discriminates between the prompt and secondary \Lcp decays.
In this fit, only candidates in the invariant mass signal region, defined to be within three times the fitted \Lcp Gaussian width of the known \Lcp mass~\cite{PDG2016} (or where a double-Gaussian signal model is used, three times the mean of widths of the two Gaussian components), are considered.
Information from the fit to the invariant mass is used to constrain the total number of \Lcp in this fit.

The shapes of the prompt and secondary \Lcp \logipchisq distributions are described by modified Novosibirsk functions~\cite{Ikeda:1999aq} with extended tail parameters.
The functional form is
\begin{equation*}
N(x;\mu;\sigma;\xi;\rhoone;\rhotwo) = 
\begin{cases}
\exp{\left[ \rhoone \frac{(x-\xone)^2}{(\mu-\xone)^2} + \frac{(\mu-\xone) \xi \sqrt{\xi^2+1} \times \sqrt{2\log{2}}}{\sigma\left(\sqrt{\xi^2 + 1} - \xi\right)^2 \log\left(\sqrt{\xi^2 + 1} +\xi \right)} - \log{2} \right]  } & x < \xone\\
\exp{\left[ -\log{2} \times \left[ \frac{\log{\left(1 + 2\xi\sqrt{\xi^2 + 1} \times \frac{x-\mu}{\sigma\sqrt{2\log{2}} }  \right)} }{\log{\left( 1 + 2\xi \left( \xi - \sqrt{\xi^2 + 1} \right) \right)  }} \right]^2 \right]   } & \xone  < x < \xtwo, \\ 
\exp{\left[ \rhotwo \frac{(\xtwo-x)^2}{(\xtwo-\mu)^2} + \frac{(\xtwo-x) \xi \sqrt{\xi^2+1} \times \sqrt{2\log{2}}}{\sigma\left(\sqrt{\xi^2 + 1} - \xi\right)^2 \log\left(\sqrt{\xi^2 + 1} +\xi \right)} - \log{2} \right]  } & x > \xtwo 
\end{cases}
\end{equation*}
where \Pxi is an asymmetry parameter, $\sigma\sqrt{2\log{2}}$ is the full-width at half maximum, \Pmu is the position of the mode, and \rhoone and \rhotwo are the lower and upper tail parameters, respectively.
The parameters \xone and \xtwo are the turnover points where the function has half of its maximum value, defined as
\begin{align*}
\xone \equiv \mu + \sigma \sqrt{2\log{2}} \left( \frac{\xi}{\sqrt{\xi^2 + 1}} - 1 \right), \\
\xtwo \equiv \mu + \sigma \sqrt{2\log{2}} \left( \frac{\xi}{\sqrt{\xi^2 + 1}} + 1 \right) .
\end{align*}
The background component is described by a nonparametric function generated using the data from the invariant mass sideband regions.
Simulated samples of prompt \Lcp and of \Lcp from a mixture of secondary \bquark-hadron decays are generated.
The values of the $\xi$, $\rhoone$, and $\rhotwo$ parameters are fixed from fits to these prompt and secondary simulated decays, while the means and widths of the functions are free to vary in the fit for the \LcpTopKmpip mode.

To aid the fit convergence in the Cabibbo-suppressed modes, where the background from unrelated tracks dominates the distribution, Gaussian constraints on the widths and means of the shapes are applied to values taken from fits to the simulation.
	The potential for bias in these shapes arising from any poor modelling of the \LcpTophh decay kinematics is investigated.
	The selection efficiency with respect to the \Lcp \logipchisq is observed to be independent of the kinematics of the \LcpTophh decays.
	The initial conditions of, and constraints applied to, the Novosibirsk shapes taken from simulation are therefore shown to be reliable.
The central value of each parameter constraint is multiplied by a scaling factor, based on the difference in the fitted value of that parameter between the data and simulated data in the unconstrained \LcpTopKmpip mode.
The fit is parameterised in terms of the prompt fraction and the total number of \Lcp candidates.
The latter has a Gaussian constraint applied to the value obtained in the fit to the \Lcp candidate invariant mass distribution.

\begin{figure}[htb]
	\captionsetup[subfigure]{labelformat=empty}	
	\centering
	\subfloat[  ]{\label{fig:promptyields:Kpi:z1}{\includegraphics[width=0.49\textwidth]{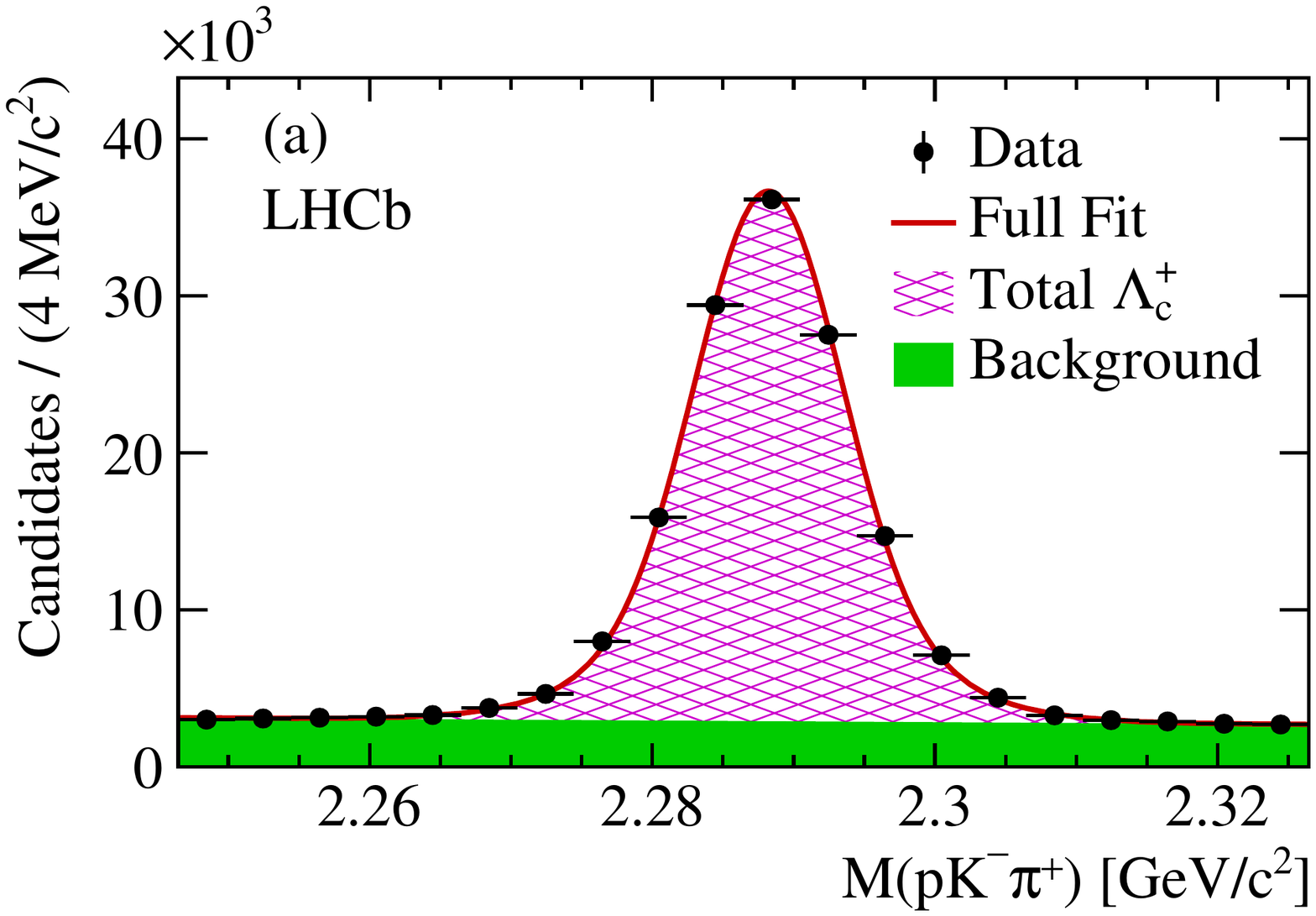}}}
	\subfloat[  ]{\label{fig:promptyields:Kpi:z2}{\includegraphics[width=0.49\textwidth]{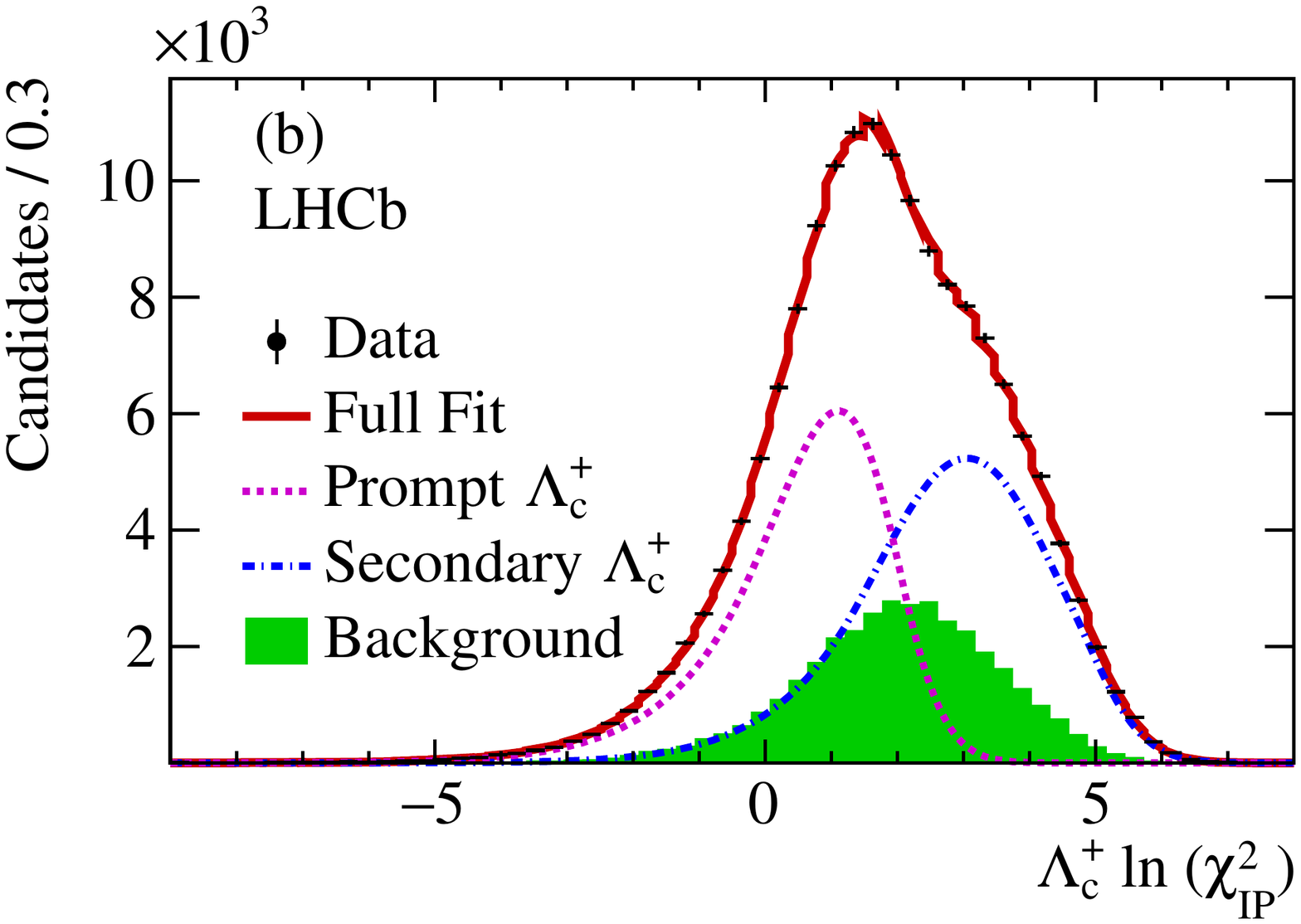}}} \\ 	\vspace{-0.9cm}
	\subfloat[  ]{\label{fig:promptyields:pipi:z1}{\includegraphics[width=0.49\textwidth]{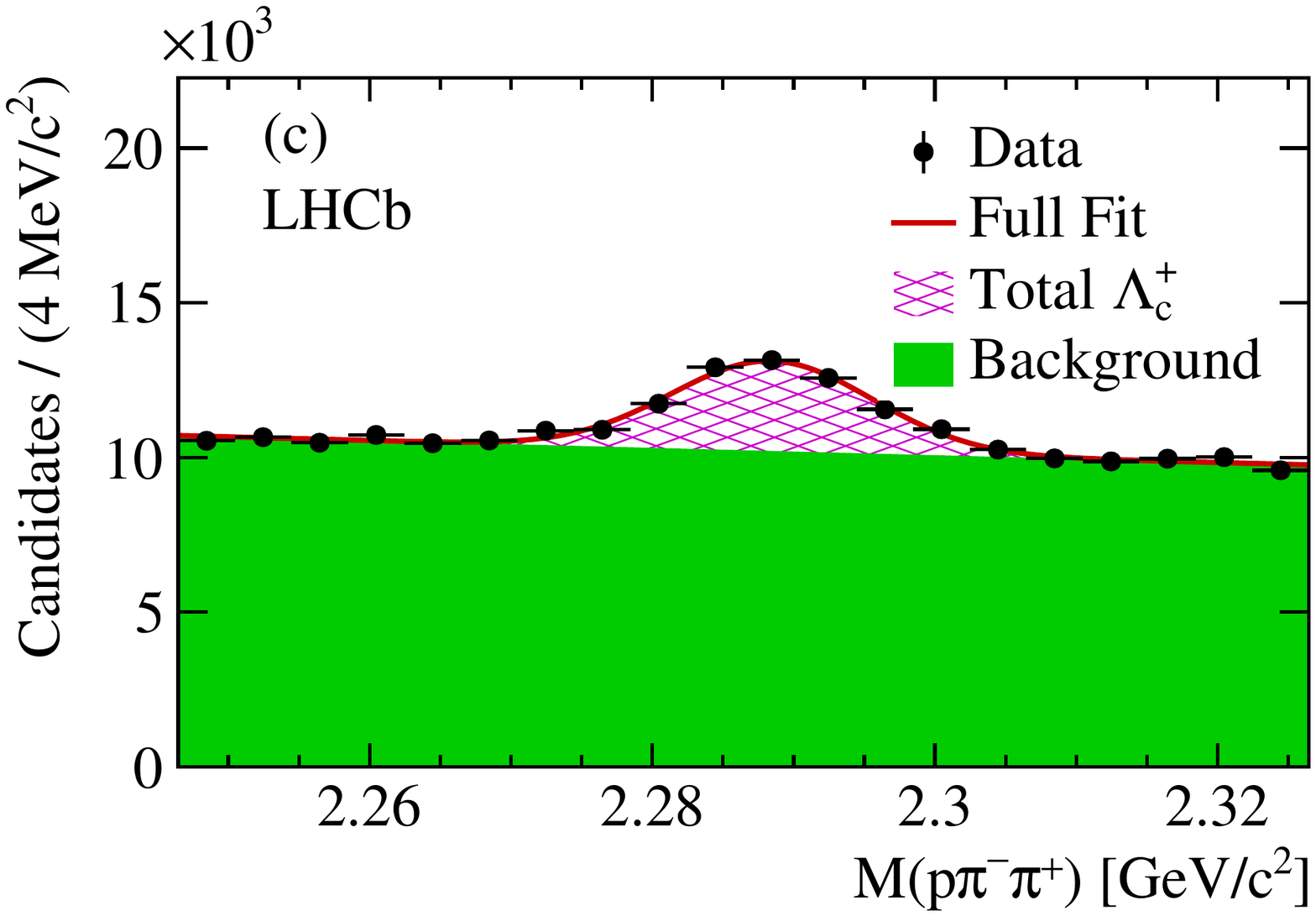}}}
	\subfloat[  ]{\label{fig:promptyields:pipi:z2}{\includegraphics[width=0.49\textwidth]{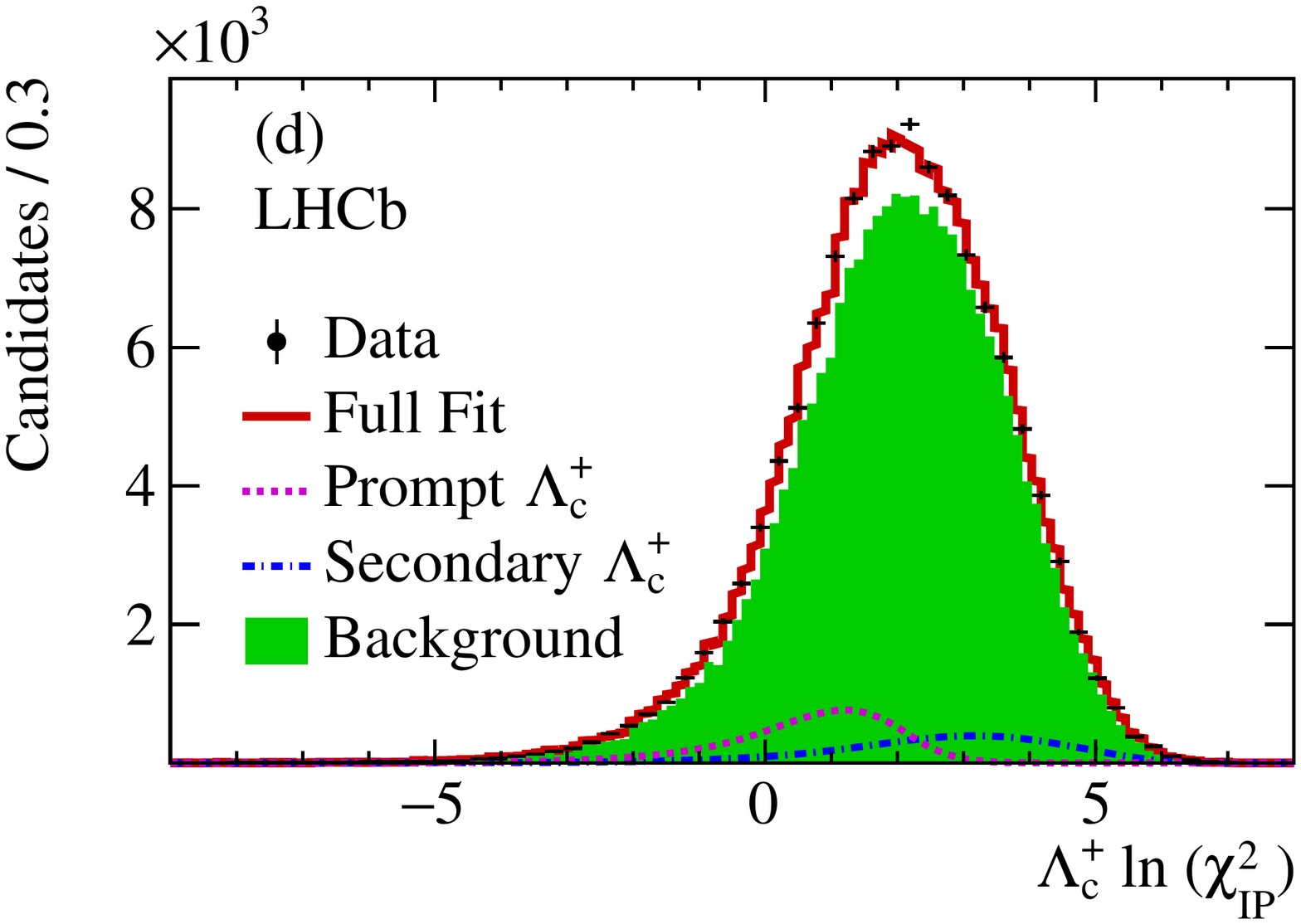}}} \\ 	\vspace{-0.9cm}
	\subfloat[  ]{\label{fig:promptyields:KK:z1}{\includegraphics[width=0.49\textwidth]{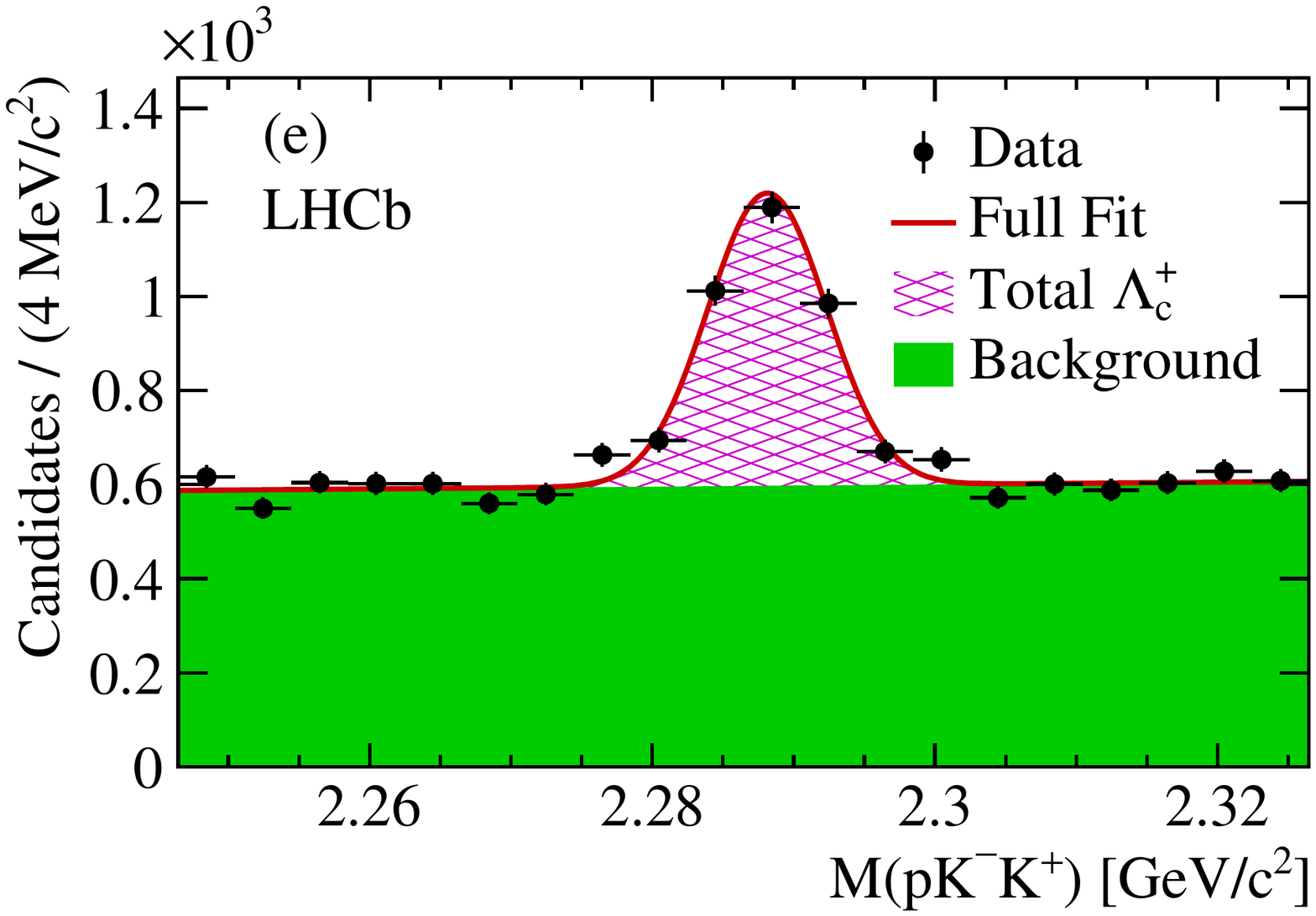}}}
	\subfloat[  ]{\label{fig:promptyields:KK:z2}{\includegraphics[width=0.49\textwidth]{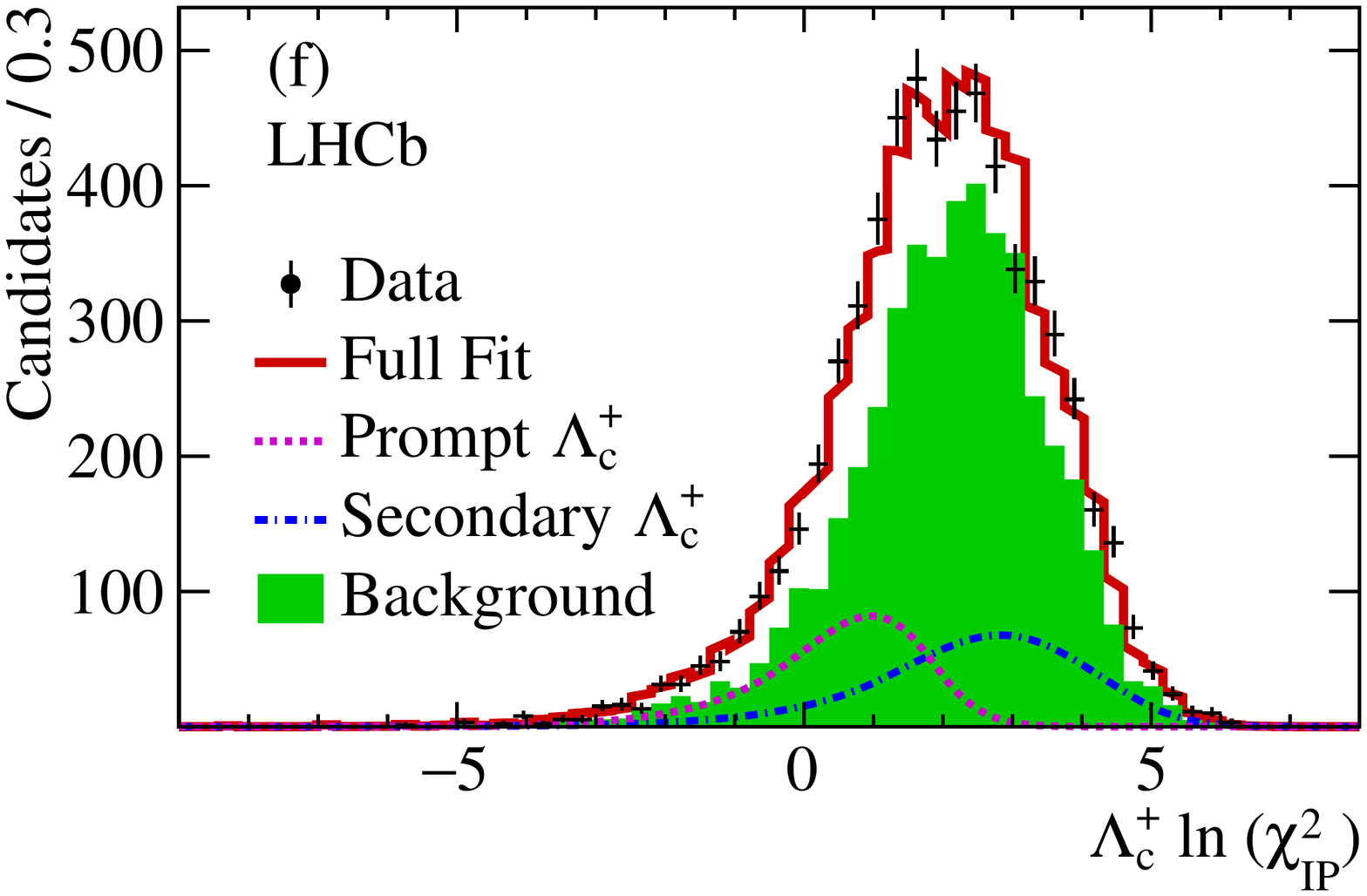}}} \\ 	\vspace{-0.74cm}
	\caption[]{
		Invariant mass distributions for (a) \LcpTopKmpip, (c) \LcpToppimpip, (e) \LcpTopKmKp in the prompt analysis, with fit results superimposed.
		The \logipchisq distributions for (b) \LcpTopKmpip, (d) \LcpToppimpip, (f) \LcpTopKmKp, with the fit results superimposed, showing the differentiation of prompt and secondary \Lcp.
		\label{fig:promptyields}}
\end{figure}

The invariant mass distributions for each of the \LcpTophh modes, with the associated fit results overlaid, are shown in the left of \Figref{fig:promptyields}, while the \logipchisq distributions and associated fit are shown on the right.
The yields in both the SL and prompt measurements are summarised in \Tabref{tab:yields}.

The fitting procedure for each decay mode is validated with a study of 1000 generated pseudoexperiments.
In each case, candidates are generated from probability density functions according to the fitted values for each decay mode, with each candidate assigned an invariant mass and a \logipchisq.
The number of candidates generated per species is the number found in the nominal fit to the data.
The fit procedure is repeated for each generated data sample as in the nominal fit.
The extracted prompt yield is shown to be unbiased, and the standard deviation on the distribution of the prompt yields verifies the reported uncertainty in the nominal fit.

\begin{table}[htp]
	\centering
		\caption[SL and prompt analysis yields]{%
		\small
		\label{tab:yields}
		Signal yields in both the SL and prompt measurements.
	}
	\begin{tabular}{l|l|r@{\ensuremath{\, \pm \,}}l}
	& \multicolumn{1}{c}{Mode} &  \multicolumn{2}{c}{Yield} \\ \hline
\multirow{4}{*}{SL}  & \LcpTopKmpip & $226{,}851$ & $522$ \\ \cline{2-4}
										& \LcpToppimpip & $19{,}584$ & $207$ \\ \cline{2-4}
										& \LcpTopKmKp & $3{,}420$ & $62$ \\ \cline{2-4}
 										& \LcpToppimKp & $392$ & $35$  \\ \hline						
\multirow{3}{*}{Prompt}  & \LcpTopKmpip & $58{,}115$ & $1{,}561$ \\ \cline{2-4}
& \LcpToppimpip & $7{,}480$ & $328$ \\ \cline{2-4}
& \LcpTopKmKp & $766$ & $61$ \\ 
\end{tabular}

\end{table}	

\section{Systematic uncertainties}
\label{sec:Syst}
Several sources of systematic uncertainty are considered in the evaluation of the selection efficiencies and in the yield determinations.
The uncertainties are summarised for the SL measurements in \Tabref{tab:syst:SL}, and for the prompt measurements in \Tabref{tab:syst:prompt}.
The systematics for the SL and prompt analyses are described together.

The uncertainties on the PID efficiencies are determined in bins of track momentum and pseudorapidity and propagated to determine the systematic uncertainties on the ratios of branching fractions.
It is assumed that the efficiency for each candidate track in a given kinematic bin is single-valued, while the finite bin size results in a kinematic distribution within each bin.
As such, small differences in the kinematic distributions of calibration and signal tracks within each bin can result in systematic errors in the assigned efficiencies.
The effect of this variation in kinematics is tested by repeating the calibration procedure with a variety of binning schemes, such that the kinematic distributions of calibration and signal tracks within each bin are altered.
After the calibration procedure has been carried out for each binning scheme and a PID selection efficiency ratio determined for each, the maximum deviation from the nominal efficiency ratio is assigned as a systematic uncertainty.
For the SL measurements this is the dominant source of systematic uncertainty, ranging from 1.4 to 2.0\,\%.

\begin{table}[htp]
	\centering
	\caption[Ratios of branching fraction systematic uncertainties - SL]{%
		\small
		\label{tab:syst:SL}
		Relative systematic uncertainties in each ratio of branching fractions, for the SL analysis.
	}
	\begin{tabular}{l|c|c|c} 
SL analysis systematic [\%]	& \LcpToppimpip	& \LcpTopKmKp & \LcpToppimKp \\ \hline
PID selection efficiency ratio	& 2.0 & 1.4 & 2.0 \\ 
Unknown \LcpTophh decay structure & 1.1 & 0.7 & 1.7 \\
Size of simulation sample  		& 0.3 & 0.3 & 0.3 \\
Trigger efficiency ratio  			& 0.6 & 0.8 & 0.3 \\
\hline 
Total												&  	2.4	&  1.8 & 2.7 \\
\end{tabular}

\end{table}	

\begin{table}[htp]
	\centering
	\caption[Ratios of branching fraction systematic uncertainties - prompt]{%
		\small
		\label{tab:syst:prompt}
		Relative systematic uncertainties in each ratio of branching fractions, for the prompt analysis.
	}
	\begin{tabular}{l|c|c} 
Prompt analysis systematic [\%]			& \LcpToppimpip	& \LcpTopKmKp \\ \cline{1-3}
PID selection efficiency ratio		& 1.2 & 1.2  \\ 
Unknown \LcpTophh decay structure & 2.7 & 3.3  \\
Yield determination uncertainty		& 3.5 & 5.7  \\
Size of simulation sample			& 0.5 & 0.5  \\ \cline{1-3}
Total													&  	4.6	&  6.7 \\
\end{tabular}

\end{table}	

The weighting procedure to align the \LcpTophh data and simulation relies upon dividing the simulation into bins of the kinematic variables describing the resonant character of the decay to evaluate the efficiency as a function of these variables.
The limited size of the simulation sample limits the precision of the description of the acceptance variation across the phase space, and therefore affects the evaluation of the selection efficiency with the weighted simulation.
Any systematic uncertainty arising from this source is evaluated through the use of generated pseudoexperiments whereby the weights assigned to the simulation in each region of the phase space are randomly resampled to determine the effect on the evaluation of selection efficiencies.
Uncertainties arising from the limited size of the simulation sample in the evaluation of the geometrical acceptance of the detector and the trigger efficiency are also assigned.

In the SL analysis imperfect modelling of variables upon which the trigger acceptance depends can lead to differences between the simulation and data which can affect the determination of the trigger acceptances.
A set of variables used in the software trigger was investigated to examine the compatibility of the data and simulation.
Where any differences were found, the simulation was reweighted individually for each variable to match the data distributions and the trigger acceptance ratios reevaluated.
A systematic uncertainty was assigned as the maximum difference between the reweighted and nominal efficiency ratios for any reweighted variable.

The systematic uncertainty on the signal yield determination is evaluated in the SL analysis by varying the choice of the fit model.
As an alternative for the signal model, a Crystal Ball function~\cite{Skwarnicki:1986xj} and a Crystal Ball function summed with a Gaussian function with a common mean are used.
The background model is modified to be a first-order or second-order polynomial.
Variations of the fit model do not result in significant changes in the signal yields and no systematic uncertainty is assigned.

For the prompt analysis the uncertainty on the determined signal yield may arise from the shape parameters that are fixed or constrained with fits to the simulated samples, and also from the limited size of the sample in the background region of the \Lcp invariant mass used to populate the background nonparametric distribution.
These are both evaluated through the use of pseudoexperiments.
The parameters governing the \logipchisq shapes are generated successively with values differing by $10\,\%$ from their fixed or constrained values in the fit; this is the maximum difference in any Novosibirsk width or mean parameter between the data and simulation fits for the \LcpTopKmpip mode, where no constraints are applied.
The background population in each bin of the template is fluctuated randomly according to a Gaussian distribution, and the fit procedure repeated.
Pseudoexperiments are also utilised to verify the statistical precision of the reported prompt \Lcp yield, and that the yields are unaffected by any bias.

The dominant systematics in the SL analysis are found to be those associated with the determination of the PID selection efficiency.
In the prompt analysis the contribution from the background template and from the constrained shape parameters are found to be the dominant uncertainties.

\section{Results}

The ratios of the branching fractions of each suppressed \LcpTophh mode relative to the \LcpTopKmpip mode are given by
\begin{equation*}
\relBR{\LcpTophh}{\LcpTopKmpip} = \frac{N(\LcpTophh) \times s_{\mathrm{scale}}}{N(\LcpTopKmpip)} \times \frac{\epsilon(\LcpTopKmpip)}{\epsilon(\LcpTophh)},
\end{equation*}
where $N$ represents the measured yield in each case, $\epsilon$ is the full selection efficiency for the mode, and $s_{\mathrm{scale}}=0.9$ is a scaling factor to account for the discarded \LcpTopKmpip data that is utilised in the selection training.
The results of the SL analysis are
\begin{equation*}
\begin{split}
\relBRpipi             & = (7.44 \pm 0.08 \pm 0.18)\,\%, \\
\relBRKK              &= (1.70 \pm 0.03 \pm 0.03)\,\%, \\
\relBRDCS & = (0.165 \pm 0.015 \pm 0.005)\,\%,
\end{split}
\end{equation*}
where the first uncertainties are statistical and the second are systematic.
Each of the measurements in the SL analysis are the most precise of these quantities to date.
In the prompt analysis the results are
\begin{equation*}
\begin{split}
\relBRpipi             & = (7.86 \pm 0.40 \pm 0.36)\,\%, \\
\relBRKK              &= (1.68 \pm 0.14 \pm 0.11)\,\%, \\
\end{split}
\end{equation*}
where the first uncertainties are statistical and the second are systematic.
The results in the prompt analysis are of comparable precision to the recent measurements at \belle~\cite{Abe:2001mb} and at BESIII~\cite{Ablikim:2016tze}.

The measurements of the ratios of the Cabibbo-suppressed branching fractions to the Cabibbo-favoured branching fraction are in agreement between the SL and prompt analyses, demonstrating that the methods employed in their determination are robust.
The efficiency correction to the ratio \relBRDCSinline is small, with the ratio of corrected and uncorrected yields differing by 3\,\%, which is comparable to the systematic uncertainty on the measurement.
The SL and prompt measurements are not combined, because the precision of such a combination would not offer a significant improvement over the precision of the SL result alone.

The measurements of the ratios of the branching fractions in the SL analysis are combined with the world-average value of the \LcpTopKmpip branching fraction, \mbox{$\mathcal{B}(\LcpTopKmpip) = (6.35 \pm 0.33)\,\%$~\cite{PDG2016}}, to compute the branching fractions of the suppressed modes
\begin{align*}
\begin{split}
\mathcal{B}(\LcpToppimpip) &= (4.72 \pm 0.05 \pm 0.11 \pm 0.25) \times 10^{-3}, \\
\mathcal{B}(\LcpTopKmKp) &= (1.08 \pm 0.02 \pm 0.02 \pm 0.06) \times 10^{-3}, \\
\mathcal{B}(\LcpToppimKp) &= (1.04 \pm 0.09 \pm 0.03 \pm 0.05) \times 10^{-4},
\end{split}
\end{align*}
where the uncertainties are statistical, systematic and due to the uncertainty of the \LcpTopKmpip branching fraction, respectively.

The measurement presented in this paper of \relBRDCSinline is lower than the value
of $(0.235\pm0.027\pm0.021)\,\%$ 
found by \belle, at the $2.0\sigma$ level,
and corresponds to $(0.58 \pm 0.06)\tan^{4}\theta_{c}$.
To account for the known flavour-SU(3) symmetry breaking that occurs due to the presence of different resonant contributions in the two modes, the fraction of the favoured decay proceeding via the $\Lambdares(1520)$ and $\Deltares^{++}$ states\footnote{Some contribution from the \W-emission decay of \decay{\Lcp}{\Deltares^0\Kp} is expected in the doubly-suppressed mode, but as argued in Ref.~\cite{PhysRevD.49.3417} the favoured decay \decay{\Lcp}{\Deltares^{++} \Km} is expected to be dominated by the \W-exchange contribution, which cannot happen in the doubly suppressed mode. The relative \W-exchange and \W-emission contributions are unknown, and the mode proceeding via a $\Deltares^{++}$ is neglected entirely.}, which cannot proceed through a doubly-suppressed transition and make up $(25 \pm 4)\, \%$ of the favoured decay, is discounted.
This yields a value of \mbox{$(0.77 \pm 0.08)\tan^{4}\theta_{c}$}.
The deviation from the naive expectation is indicative that either \mbox{\W-exchange} contributions to the favoured mode are more significant than previously believed, or that some flavour-SU(3) symmetry breaking effect not present in the charmed-meson sector is present in the charmed-baryon sector, or some combination of the two.

Future analysis of the resonant character of the \LcpTophh decays, through which such symmetry breaking effects occur will be important in establishing the nature of this effect.
In particular the comparison of individual resonant contributions which can proceed through \W-exchange in the favoured mode but not the doubly suppressed mode, such as \decay{\Lcp}{\Deltares^{++} \Km}, \decay{\Deltares^{++}}{\proton \pip} and \decay{\Lcp}{\Deltares^0 \Kp}, \decay{\Deltares^0}{\proton \pim}, will provide a stronger statement about the prominence of \W-exchange diagrams in the charmed-baryon sector.

\section*{Acknowledgements}

\noindent We express our gratitude to our colleagues in the CERN
accelerator departments for the excellent performance of the LHC. We
thank the technical and administrative staff at the LHCb
institutes. We acknowledge support from CERN and from the national
agencies: CAPES, CNPq, FAPERJ and FINEP (Brazil); MOST and NSFC
(China); CNRS/IN2P3 (France); BMBF, DFG and MPG (Germany); INFN
(Italy); NWO (The Netherlands); MNiSW and NCN (Poland); MEN/IFA
(Romania); MinES and FASO (Russia); MinECo (Spain); SNSF and SER
(Switzerland); NASU (Ukraine); STFC (United Kingdom); NSF (USA).  We
acknowledge the computing resources that are provided by CERN, IN2P3
(France), KIT and DESY (Germany), INFN (Italy), SURF (The
Netherlands), PIC (Spain), GridPP (United Kingdom), RRCKI and Yandex
LLC (Russia), CSCS (Switzerland), IFIN-HH (Romania), CBPF (Brazil),
PL-GRID (Poland) and OSC (USA). We are indebted to the communities
behind the multiple open-source software packages on which we depend.
Individual groups or members have received support from AvH Foundation
(Germany), EPLANET, Marie Sk\l{}odowska-Curie Actions and ERC
(European Union), ANR, Labex P2IO, ENIGMASS and OCEVU, and R\'{e}gion
Auvergne-Rh\^{o}ne-Alpes (France), RFBR and Yandex LLC (Russia), GVA,
XuntaGal and GENCAT (Spain), Herchel Smith Fund, the Royal Society,
the English-Speaking Union and the Leverhulme Trust (United Kingdom).

\addcontentsline{toc}{section}{References}
\setboolean{inbibliography}{true}
\bibliographystyle{LHCb}
\bibliography{supp,LHCb-PAPER,LHCb-CONF,LHCb-DP,LHCb-TDR,myrefs}

\newpage
\centerline{\large\bf LHCb collaboration}
\begin{flushleft}
\small
R.~Aaij$^{40}$,
B.~Adeva$^{39}$,
M.~Adinolfi$^{48}$,
Z.~Ajaltouni$^{5}$,
S.~Akar$^{59}$,
J.~Albrecht$^{10}$,
F.~Alessio$^{40}$,
M.~Alexander$^{53}$,
A.~Alfonso~Albero$^{38}$,
S.~Ali$^{43}$
% ...

{\footnotesize \it
% ...
$ ^{5}$Clermont Universit{\'e}, Universit{\'e} Blaise Pascal, CNRS/IN2P3, LPC, Clermont-Ferrand, France\\
% ...
$ ^{10}$Fakult{\"a}t Physik, Technische Universit{\"a}t Dortmund, Dortmund, Germany\\
% ...
$ ^{38}$ICCUB, Universitat de Barcelona, Barcelona, Spain\\
$ ^{39}$Universidad de Santiago de Compostela, Santiago de Compostela, Spain\\
$ ^{40}$European Organization for Nuclear Research (CERN), Geneva, Switzerland\\
% ...
$ ^{43}$Nikhef National Institute for Subatomic Physics, Amsterdam, The Netherlands\\
% ...
$ ^{48}$H.H. Wills Physics Laboratory, University of Bristol, Bristol, United Kingdom\\
% ...
$ ^{53}$School of Physics and Astronomy, University of Glasgow, Glasgow, United Kingdom\\
% ...
$ ^{59}$University of Cincinnati, Cincinnati, OH, United States\\
% ...
}
\end{flushleft}

\end{document}